\documentclass[prd,aps,twocolumn,eqsecnum,nofootinbib]{revtex4-2}

\usepackage{amsmath,amssymb,amsfonts}

\usepackage{graphicx}
\usepackage{adjustbox}
\usepackage{dcolumn}
\usepackage{bm}
\usepackage{longtable}
\usepackage[table]{xcolor}
\usepackage{diagbox}
\usepackage{multirow}
\usepackage{hhline}
\usepackage{xstring}
\usepackage{etoolbox}
\usepackage{orcidlink}

\usepackage{hyperref}
\hypersetup{
  colorlinks = true,
  linkcolor  = blue,
  citecolor  = blue,
  urlcolor   = blue
}

\begin{document}


\title{Letelier black hole immersed in an electromagnetic universe}

\author{Ahmad Al-Badawi\orcidlink{0000-0002-3127-3453}}
\email{ahmadbadawi@ahu.edu.jo}
\affiliation{Department of Physics, Al-Hussein Bin Talal University, Ma'an 71111, Jordan}

\author{Faizuddin Ahmed\orcidlink{0000-0003-2196-9622}}
\email{faizuddinahmed15@gmail.com}
\affiliation{Department of Physics, The Assam Royal Global University, Guwahati 781035, Assam, India}

\author{\.{I}zzet Sakall{\i}\orcidlink{0000-0001-7827-9476}}
\email{izzet.sakalli@emu.edu.tr (Corresp. author)}
\affiliation{Department of Physics, Eastern Mediterranean University, Famagusta Northern Cyprus 99628, via Mersin 10, Turkiye}

\date{\today}

\begin{abstract}
We investigate a static, spherically symmetric black hole solution surrounded by a cloud of strings and immersed in an electromagnetic universe. By deriving the event horizon from the lapse function, we demonstrate that both the string cloud parameter and the electromagnetic background parameter significantly modify the horizon radius compared to the Schwarzschild case. Consequently, thermodynamic quantities-including the Hawking temperature, Bekenstein-Hawking entropy, and heat capacity-become explicit functions of these additional parameters, with the heat capacity exhibiting divergences that signal phase transitions. We analyze the motion of massive test particles in this spacetime, deriving the effective potential and calculating the innermost stable circular orbit radius, which governs the inner edge of accretion disks and influences orbital stability. Scalar perturbations are examined through the associated effective potential, and quasinormal mode frequencies are computed using the sixth-order WKB approximation; the negative imaginary parts confirm the stability of the black hole under such perturbations. We also study the photon sphere structure, black hole shadow radius, and photon trajectories, showing how the interplay between string clouds and the electromagnetic background shapes the optical properties of this spacetime. Finally, we investigate weak gravitational lensing phenomena by deriving the deflection angle for both massive particles and photons using the Gauss-Bonnet theorem applied to the optical geometry. The results exhibit notable deviations from the Schwarzschild geometry, with the string cloud enhancing the deflection through a $(1-\alpha)^{-1}$ factor while the electromagnetic parameter introduces competing corrections at second order.
\end{abstract}

\maketitle

\small

\section{Introduction} \label{isec1}

Black holes (BHs) stand among the most fascinating predictions of Einstein's general relativity, serving as natural laboratories where gravitational effects reach their extreme limits \cite{isz01,isz02}. Over the past decades, direct observational evidence from gravitational wave detections by LIGO/Virgo \cite{isz03,isz04} and the shadow imaging by the Event Horizon Telescope \cite{isz05,isz06} has transformed BH physics from a theoretical curiosity into an observationally grounded discipline. These developments have intensified interest in understanding how BH properties are modified when surrounded by various matter distributions or immersed in external fields \cite{isz07,isz08}.

In realistic astrophysical environments, BHs rarely exist in isolation. They are typically surrounded by accretion disks, dark matter halos, magnetic fields, or more exotic matter configurations that can leave observable imprints on their spacetime geometry \cite{isz09,isz10}. Among such configurations, the cloud of strings (CoS) introduced by Letelier \cite{isz11} has attracted considerable attention. Letelier demonstrated that a spherically symmetric distribution of strings threading spacetime produces an energy-momentum tensor with $T^t_t = T^r_r = \alpha/r^2$ and vanishing angular components, where $\alpha$ denotes the string cloud parameter. The resulting Letelier BH solution modifies the Schwarzschild metric through a global deficit factor $(1-\alpha)$, which alters the horizon structure, thermodynamic quantities, and geodesic motion \cite{isz12,isz13,isz14,Aydiner:2025eii}. Cosmic strings themselves are one-dimensional topological defects predicted by grand unified theories and may have formed during phase transitions in the early universe \cite{isz15,isz16}. Their gravitational signatures, including lensing effects and contributions to the cosmic microwave background, remain active areas of investigation \cite{isz17,isz18}. BH solutions in a wide range of gravity theories, when coupled to a surrounding cloud of strings, have attracted substantial attention in recent years. Such configurations provide a useful framework for exploring how string-like matter distributions modify spacetime geometry and affect the physical observables associated with BHs. In particular, the presence of a CoS can introduce angular deficits, alter horizon structure, and influence geodesic motion as well as lensing and shadow characteristics. These effects become even more intriguing when the BH is further immersed in external fields-such as dark matter halos, quintessence, nonlinear electrodynamics, or higher-curvature corrections-for representative examples, see, for instance \cite{epjc1,epjc2,epjc3,Sucu:2025olo,Gursel:2025wan} and references therein.

Parallel to these developments, BH solutions embedded in external electromagnetic (EM) fields have been studied extensively. The Bertotti-Robinson (BR) spacetime \cite{isz19,isz20} describes a conformally flat geometry supported by a uniform EM field and represents the near-horizon limit of extremal Reissner-Nordstr\"om (RN) BHs. Building upon this foundation, Halilsoy and collaborators \cite{isz221,Halilsoy1998} constructed the Schwarzschild electromagnetic BH (SEBH) solution, which interpolates between the Schwarzschild BH (SBH) and the extremal RN geometry through an electromagnetic universe (EMU) parameter $a$ satisfying $0 < a \leq 1$. When $a=1$, the standard SBH is recovered, while $a \to 0$ approaches the extremal RN limit transformable to the BR metric. This interpolation scheme provides a useful framework for examining how external EM backgrounds influence BH physics \cite{Halilsoy1995,isz22,Ovgun2016,isz23,isz24,badaw1}.

The thermodynamics of BHs has been a cornerstone of theoretical physics since Bekenstein's entropy formula \cite{isz25} and Hawking's discovery of thermal radiation \cite{isz26}. For modified BH solutions, quantities such as the Hawking temperature, entropy, and heat capacity acquire corrections that encode information about the surrounding matter or fields \cite{isz27,isz28,isz29}. Phase transitions, signaled by divergences in the heat capacity, reveal critical behavior analogous to ordinary thermodynamic systems \cite{isz30,isz31}. Understanding how CoS and EMU parameters affect these thermodynamic properties offers potential observational discriminants between different BH models.

The motion of test particles around BHs provides another diagnostic tool. The innermost stable circular orbit (ISCO) plays a central role in accretion disk physics, determining the inner edge of the disk and influencing the observed spectrum \cite{isz32,isz33}. Modifications to the ISCO radius by exotic matter or external fields can thus have direct observational consequences. Similarly, photon orbits define the BH shadow boundary, which has become directly measurable through very long baseline interferometry \cite{isz34,isz35}.

Quasinormal modes (QNMs) characterize the damped oscillations of perturbed BHs and carry unique fingerprints of the underlying spacetime geometry \cite{isz36,isz37}. The complex QNM frequencies, with real parts representing oscillation rates and imaginary parts representing damping times, depend solely on BH parameters and are therefore valuable for testing gravity theories. The WKB approximation, developed by Schutz, Will, Iyer, and Konoplya \cite{isz38,isz39,isz40}, provides an efficient semi-analytical method for computing these frequencies to high accuracy.

Gravitational lensing by BHs offers yet another observational window. The deflection of light rays passing near a BH depends sensitively on the spacetime geometry, and measurements of lensing angles can constrain BH parameters \cite{isz41,isz42,Sucu:2025fwa,Sucu:2025lqa}. The Gauss-Bonnet theorem applied to the optical geometry provides an elegant framework for calculating deflection angles in static, spherically symmetric spacetimes \cite{isz43,isz44}.

Despite the extensive literature on Letelier BH and SEBH solutions separately, the combined system-a Letelier BH immersed in an EMU-has not been thoroughly examined. This configuration is physically motivated: cosmic strings in the early universe would have existed alongside primordial magnetic fields, and their combined gravitational effects merit investigation. The interplay between the global geometric modification from CoS (through the $\alpha$ parameter) and the local metric correction from the EMU (through the $a$ parameter) produces a rich phenomenology that differs qualitatively from either effect alone.

In this work, we construct and analyze the Letelier BH solution immersed in EMU. Our primary objectives are: (i) to determine the horizon structure and identify the parameter regimes yielding non-extremal, extremal, and naked singularity configurations; (ii) to compute thermodynamic quantities including the Hawking temperature, entropy, and heat capacity, and to locate phase transitions; (iii) to study timelike geodesics, derive the effective potential, and calculate the ISCO radius as functions of $\alpha$ and $a$; (iv) to analyze null geodesics, obtain the photon sphere radius, and determine the shadow size; (v) to investigate scalar perturbations and compute QNM frequencies using the sixth-order WKB method; and (vi) to derive the weak-field gravitational deflection angle for both massive particles and photons using the Gauss-Bonnet approach. Through these investigations, we aim to elucidate how the presence of cosmic string matter and an EM background jointly modify observable BH properties.

The paper is organized as follows. In Sec.~\ref{isec2}, we present the spacetime metric for the Letelier BH immersed in EMU, derive the horizon structure, and discuss limiting cases. Section~\ref{isec3} examines the remarkable physical features of this BH, beginning with thermodynamics in Sec.~\ref{isec3}A, followed by test particle dynamics in Sec.~\ref{isec3}B, scalar perturbations and QNMs in Secs.~\ref{isec3}C--D, and photon sphere properties in Sec.~\ref{isec3}E. Section~\ref{sec4} is devoted to weak gravitational lensing, where we derive the deflection angle and analyze its behavior across parameter space. Finally, Sec.~\ref{sec5} summarizes our findings and discusses potential extensions of this work.

\section{Spacetime of Letelier BH Immersed in EMU} \label{isec2}

A static and spherically symmetric SBH immersed in an EMU is described by the following line element \cite{isz221,Halilsoy1998}:
\begin{equation}
ds^{2}=-A(r)\, dt^{2}+\frac{dr^{2}}{A(r)}+r^{2}\left(d\theta ^{2}+\sin ^{2}\theta\, d\phi ^{2}\right), \label{aa1}
\end{equation}
with the lapse function
\begin{equation}
A(r)=1-\frac{2M}{r}+\frac{M^{2}}{r^{2}}\left( 1-a^{2}\right). \label{aa1b}
\end{equation}
Here $M$ denotes the BH mass coupled to an external EM field, and $a$ $(0<a\leq 1)$ is the interpolation parameter, commonly referred to as the EMU parameter \cite{isz19,isz20}. This metric satisfies all required boundary conditions: when $a=1$, we recover the standard SBH metric, while for $0<a<1$, we obtain the SEBH solution. The limiting case $a=0$, which is excluded from this family, corresponds to the extremal RN geometry that can be transformed into the BR metric \cite{sec2is01,sec2is02}. The metric~(\ref{aa1}) admits two horizons located at
\begin{equation}
r_{-}=M\left( 1-a\right),\qquad r_{+}=M\left( 1+a\right). \label{aa1c}
\end{equation}
The outer horizon $r_{+}$ reduces to the event horizon $r=2M$ of the SBH when $a=1$, while the inner horizon $r_{-}$ vanishes in this limit.

Our goal is to investigate the above BH solution when surrounded by a CoS. The original formulation for incorporating a CoS as a gravitational source within Einstein's theory was developed by Letelier \cite{isz11}, who established a consistent framework for describing such extended one-dimensional objects in curved spacetime. Cosmic strings are topological defects that may have formed during symmetry-breaking phase transitions in the early universe, and their gravitational effects have been studied in various astrophysical and cosmological contexts \cite{sec2is03,sec2is04}. Using Letelier's approach, one obtains a generalized SBH solution representing a BH surrounded by a spherically symmetric distribution of strings. In this model, the presence of the CoS modifies the spacetime geometry through an anisotropic energy-momentum tensor whose nonvanishing components are \cite{isz11}:
\begin{equation}
    T^{t\,(\rm CS)}_{t}=\rho^{\rm CS}=\frac{\alpha}{r^2}=T^{r\,(\rm CS)}_{r},\quad T^{\theta\,(\rm CS)}_{\theta}=T^{\phi\,(\rm CS)}_{\phi}= 0,\label{aa3}
\end{equation}
where $\rho^{\mathrm{CS}}$ denotes the energy density associated with the CoS and $\alpha$ is an integration constant characterizing the strength of the string distribution. This particular form of the energy-momentum tensor reflects the fact that CoS matter exerts radial pressure while contributing no tangential stresses, thereby distinguishing it from conventional perfect fluids. The resulting spacetime exhibits deviations from the vacuum SBH geometry, encoding the gravitational influence of the string cloud and providing a foundation for studying how such extended objects affect BH structure, orbital dynamics, perturbations, and observational signatures \cite{sec2is05}.

We now consider a static, spherically symmetric spacetime in the presence of a CoS immersed in an EMU, described in standard spherical coordinates $(t, r, \theta, \phi)$. This setup allows us to investigate the combined effects of the CoS and EM background on the geometry and dynamics of the spacetime. The line element takes the form
\begin{equation}
ds^{2}=-f\left( r\right) dt^{2}+\frac{dr^{2}}{f\left( r\right) }+r^{2}\left(
d\theta ^{2}+\sin ^{2}\theta\, d\phi ^{2}\right), \label{aa4}
\end{equation}
where the metric function is given by
\begin{equation}
f\left( r\right) =1-\alpha-\tfrac{2M}{r}+\tfrac{M^{2}}{r^{2}}\left( 1-a^{2}\right).\label{aa5}
\end{equation}
Here the CoS parameter $\alpha$ lies in the range $0 \leq \alpha <1$, while $a$ is the EMU parameter defined earlier. The metric function~(\ref{aa5}) reduces to several well-known cases: (i) when $\alpha=0$ and $a=1$, we recover the SBH; (ii) when $\alpha=0$ and $0<a<1$, we obtain the SEBH; (iii) when $\alpha\neq 0$ and $a=1$, we get the Letelier BH with CoS; and (iv) when $\alpha=0$ and $a\to 0$, we approach the extremal RN limit.

The horizon structure is determined by the condition
\begin{equation}
    f(r_{h})=0\quad\Rightarrow\quad 1-\alpha-\tfrac{2M}{r_{h}}+\tfrac{(1-a^2) M^{2}}{r^{2}_{h}}=0,\label{condition}
\end{equation}
which yields the horizon radii
\begin{equation}
r_{\pm}=\tfrac{M}{1-\alpha}\left(1\pm\sqrt{1-(1-\alpha) (1-a^2)}\right).\label{aa7}
\end{equation}
For these radii to be real and positive, the discriminant must satisfy $(1-\alpha)(1-a^2)\leq 1$. When $(1-\alpha)(1-a^2)=1$, the two horizons merge into a single degenerate horizon, yielding an extremal BH configuration. When $(1-\alpha)(1-a^2)>1$, no real horizons exist, and the spacetime contains a naked singularity. This condition can be rewritten as $\alpha \geq 1-(1-a^2)^{-1}$, which for $a=1$ (SBH limit) becomes $\alpha\geq 1$—hence the SBH with CoS develops a naked singularity when $\alpha$ approaches unity from below.

Table~\ref{izzetletelier_horizons} presents the horizon structure for various combinations of the EMU parameter $a$ and the CoS parameter $\alpha$. Several notable features emerge from this data. First, the extremal RN limit ($a=0$, $\alpha=0$) exhibits a single degenerate horizon at $r_h=M$; introducing even a small CoS density ($\alpha=0.01$) immediately splits this into distinct inner and outer horizons at $r_-=0.9091M$ and $r_+=1.1111M$. Second, the SBH limit ($a=1$, $\alpha=0$) shows a single horizon at $r_h=2M$, which shifts outward as $\alpha$ increases: $r_h=2.02M$ for $\alpha=0.01$, reaching $r_h=4M$ for $\alpha=0.5$, before naked singularity formation occurs at $\alpha=0.9$. Third, for intermediate EMU parameter values ($0<a<1$), the double-horizon structure persists across all $\alpha$ values until extremality is reached near $\alpha \approx 0.9$. Fourth, the inner horizon location decreases with increasing $a$, following approximately $r_- \approx M(1-a)/(1-\alpha)$, while the outer horizon increases with both $\alpha$ and $a$.

\setlength{\tabcolsep}{12pt}
\renewcommand{\arraystretch}{1.6}
\arrayrulecolor{black}

\begin{longtable*}{|c|c|c|c|}
\caption{Horizon structure of Letelier black holes with a cloud of strings (CoS) in the EMU background for selected combinations of the EMU parameter $a$ and CoS parameter $\alpha$ at fixed mass $M=1$. The table shows that: (i) in the extremal RN limit ($a=0$, $\alpha=0$) there is a single degenerate horizon at $r_h = 1.0M$, which splits into inner and outer horizons once the CoS is switched on; (ii) in the Schwarzschild limit ($a=1$, $\alpha=0$) the single horizon at $r_h = 2M$ shifts outward with increasing $\alpha$ until a naked singularity appears at $\alpha = 0.9$; (iii) for $0<a<1$ a double-horizon structure persists up to near-extremal configurations around $\alpha \simeq 0.9$; (iv) the outer horizon grows with both $\alpha$ and $a$, while the inner horizon approximately follows $r_- \simeq M(1-a)/(1-\alpha)$.}
\label{izzetletelier_horizons}\\
\hline
\rowcolor{orange!50}
\textbf{$a$} & \textbf{$\alpha$} & \textbf{Horizon(s)} & \textbf{Configuration} \\
\hline
\endfirsthead

\hline
\rowcolor{orange!50}
\textbf{$a$} & \textbf{$\alpha$} & \textbf{Horizon(s)} & \textbf{Configuration} \\
\hline
\endhead

0.0 & 0.0 & $[1.0]$                  & Extremal BH \\
\hline
0.0 & 0.01 & $[0.9091,\ 1.1111]$     & Non-extremal (inner+outer) \\
\hline
0.0 & 0.1 & $[0.7597,\ 1.4625]$      & Non-extremal (inner+outer) \\
\hline
0.0 & 0.5 & $[0.5858,\ 3.4142]$      & Non-extremal (inner+outer) \\
\hline
0.0 & 0.9 & $[0.5132]$               & Extremal BH \\
\hline
0.1 & 0.0 & $[0.9000,\ 1.1000]$      & Non-extremal (inner+outer) \\
\hline
0.1 & 0.01 & $[0.8676,\ 1.1526]$     & Non-extremal (inner+outer) \\
\hline
0.1 & 0.1 & $[0.7443,\ 1.4779]$      & Non-extremal (inner+outer) \\
\hline
0.1 & 0.5 & $[0.5787,\ 3.4213]$      & Non-extremal (inner+outer) \\
\hline
0.1 & 0.9 & $[0.5079]$               & Extremal BH \\
\hline
0.5 & 0.0 & $[0.5000,\ 1.5000]$      & Non-extremal (inner+outer) \\
\hline
0.5 & 0.01 & $[0.4975,\ 1.5227]$     & Non-extremal (inner+outer) \\
\hline
0.5 & 0.1 & $[0.4777,\ 1.7445]$      & Non-extremal (inner+outer) \\
\hline
0.5 & 0.5 & $[0.4189,\ 3.5811]$      & Non-extremal (inner+outer) \\
\hline
0.5 & 0.9 & $[0.3823]$               & Extremal BH \\
\hline
1.0 & 0.0 & $[2.0]$                  & SBH \\
\hline
1.0 & 0.01 & $[2.0202]$              & Letelier BH \\
\hline
1.0 & 0.1 & $[2.2222]$               & Letelier BH \\
\hline
1.0 & 0.5 & $[4.0]$                  & Letelier BH \\
\hline
1.0 & 0.9 & $[\,]$                   & Naked singularity \\
\hline
\end{longtable*}

\begin{figure*}[!ht]
    \centering
    \includegraphics[width=0.96\textwidth]{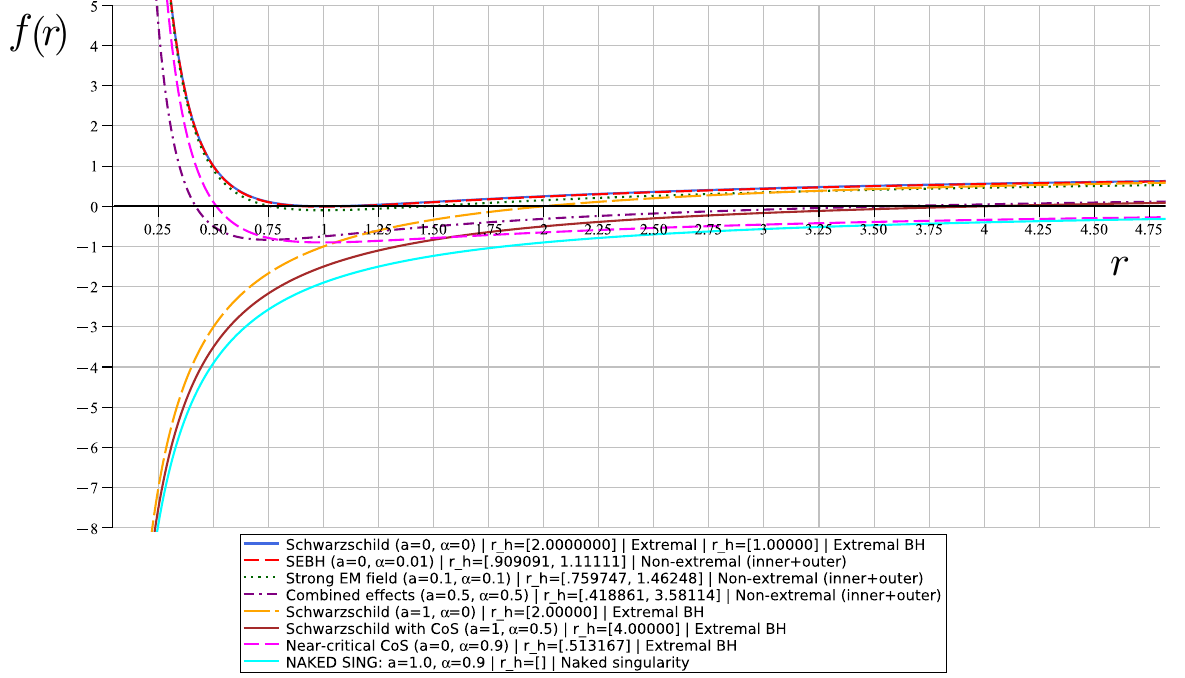}
    \caption{\footnotesize Metric function $f(r)$ for Letelier BHs with CoS immersed in EMU with fixed mass $M=1$. Horizons occur where $f(r_h) = 0$ (black horizontal line). Eight distinct configurations are shown: (i) \textcolor{blue}{Blue solid}: extremal RN limit ($a=0$, $\alpha=0$) with degenerate horizon at $r_h = 1.0M$; (ii) \textcolor{red}{Red dashed}: SEBH with minimal CoS ($a=0$, $\alpha=0.01$) showing split horizons at $r_h = [0.909M, 1.111M]$; (iii) \textcolor{black}{Black dotted}: strong EM field with moderate CoS ($a=0.1$, $\alpha=0.1$) yielding $r_h = [0.760M, 1.462M]$; (iv) \textcolor{purple}{Purple dash-dot}: combined strong effects ($a=0.5$, $\alpha=0.5$) with widely separated horizons at $r_h = [0.419M, 3.581M]$; (v) \textcolor{orange}{Orange solid}: SBH limit ($a=1$, $\alpha=0$) with $r_h = 2.0M$; (vi) \textcolor{brown}{Brown solid}: SBH with moderate CoS ($a=1$, $\alpha=0.5$) pushing the horizon to $r_h = 4.0M$; (vii) \textcolor{magenta}{Magenta solid}: near-critical CoS ($a=0$, $\alpha=0.9$) with extremal configuration at $r_h = 0.513M$; (viii) \textcolor{cyan}{Cyan solid}: naked singularity case ($a=1.0$, $\alpha=0.9$) with no horizons. All curves approach $f(r) \to 1 - \alpha$ as $r \to \infty$, reflecting the global effect of CoS on the spacetime geometry.}
    \label{izzetletelier_metric_function}
\end{figure*}

The quantitative horizon data in Table~\ref{izzetletelier_horizons} confirms the qualitative trends displayed in Fig.~\ref{izzetletelier_metric_function}. The horizon separation $\Delta r = r_+ - r_-$ increases superlinearly with the CoS parameter $\alpha$: at $a=0.5$, the separation grows from $\Delta r = 1.0M$ at $\alpha=0$ to $\Delta r = 3.16M$ at $\alpha=0.5$, representing more than threefold expansion. Moreover, Table~\ref{izzetletelier_horizons} demonstrates the critical threshold behavior at $\alpha \approx 0.9$, where all non-SBH configurations ($a<1$) transition to extremality with merged horizons, while the SBH case ($a=1$) violates cosmic censorship entirely, producing a naked singularity at this same critical value.

Figure~\ref{izzetletelier_metric_function} displays the metric function $f(r)$ for various parameter combinations, revealing the interpolation between EMU effects and CoS matter. The horizontal zero line serves as the critical diagnostic for horizon locations, where each curve's intersection marks either event horizons or inner horizons depending on the configuration. Several physical features emerge from this comparison.

First, the extremal RN configuration (blue solid curve, $a=0$, $\alpha=0$) represents the baseline case with a single degenerate horizon at $r_h = M = 1.0$, corresponding to the BR metric limit that is deliberately excluded from the SEBH family due to its special conformal structure. Introducing even minimal CoS density ($\alpha=0.01$, red dashed curve) immediately breaks this degeneracy, splitting the horizon into distinct inner and outer components at $r_h = [0.9091M, 1.1111M]$. This demonstrates the sensitivity of the extremal RN configuration to matter perturbations. The metric function exhibits characteristic behavior near these horizons: a local maximum between the two roots indicates a region where the timelike Killing vector becomes spacelike.

Second, as we increase both the EMU parameter and CoS density (black dotted curve, $a=0.1$, $\alpha=0.1$), the horizon separation widens dramatically, with the inner horizon moving inward to $r_- = 0.7443M$ while the outer horizon expands to $r_+ = 1.4779M$. This trend continues with the purple dash-dot curve ($a=0.5$, $\alpha=0.5$), where strong CoS pushes the outer horizon to $r_+ = 3.5811M$ while compressing the inner horizon to $r_- = 0.4189M$. The deep negative well between these horizons, extending to $f(r) \approx -0.7$, reflects the strong gravitational potential created by the EM field competing with the cosmic string tension.

Third, the SBH limit cases (orange and brown curves, $a=1$) provide crucial reference points: without CoS ($\alpha=0$), we recover the classical SBH horizon at $r_h = 2M$ as expected. Adding CoS ($\alpha=0.5$, brown curve) shifts this single horizon outward to $r_h = 4M$, demonstrating that CoS matter reduces the effective gravitational strength by contributing negative radial pressure. The formula $f(r) = 1 - \alpha - 2M/r$ in the SBH limit ($a=1$) makes this effect transparent: CoS effectively reduces the lapse function by a constant $\alpha$, requiring larger radii to achieve $f(r_h) = 0$.

Fourth, the asymptotic behavior visible in Fig.~\ref{izzetletelier_metric_function} is particularly instructive: all curves approach the constant value $f(r) \to 1 - \alpha$ as $r \to \infty$, rather than unity as in pure SBH geometry. This persistent offset reflects the global geometric effect of the CoS, which maintains a uniform energy density $\rho^{\text{CS}} = \alpha/r^2$ throughout space. The departure from asymptotic flatness becomes increasingly pronounced for larger $\alpha$ values, with the $\alpha=0.5$ cases approaching $f(\infty) = 0.5$, indicating that half the asymptotic ``rest energy'' is consumed by the cosmic string network.

Fifth, comparing curves with fixed $\alpha$ but varying $a$ reveals the EMU effect: smaller $a$ values (stronger external EM fields) produce tighter double-horizon configurations clustered closer to $r \sim M$, while $a \to 1$ (SBH limit) eliminates the inner horizon entirely, leaving only the event horizon. Conversely, fixing $a$ and increasing $\alpha$ demonstrates the CoS pushing both horizons: the inner horizon moves slowly inward while the outer horizon accelerates dramatically outward, with the separation $\Delta r = r_+ - r_-$ growing nonlinearly with $\alpha$.

Table~\ref{tab:1} provides a parametric survey of the dimensionless outer horizon radius $r_+/M$ across the $(a,\alpha)$ parameter space. The data confirms that $r_+$ increases monotonically with both parameters. For the SBH limit ($a=1$), the horizon radius follows $r_+/M = 2/(1-\alpha)$, yielding $r_+=2.0M$ at $\alpha=0$, $r_+=2.222M$ at $\alpha=0.1$, and $r_+=2.5M$ at $\alpha=0.2$. For the extremal RN limit ($a=0$), smaller horizon values are obtained, ranging from $r_+=1.0M$ at $\alpha=0$ to $r_+=1.809M$ at $\alpha=0.2$. The intermediate EMU parameter values interpolate smoothly between these extremes.

\setlength{\tabcolsep}{12pt}
\renewcommand{\arraystretch}{1.6}
\arrayrulecolor{black}

\begin{longtable*}{|>{\columncolor{orange!50}}c|c|c|c|c|c|c|}
\caption{\footnotesize Dimensionless outer horizon radius $r_{+}/M$ for various combinations of the CoS parameter $\alpha$ and EMU parameter $a$. The horizon radius grows monotonically with both parameters, ranging from $r_{+}=1.0M$ at the extremal RN limit ($a=0$, $\alpha=0$) up to $r_{+}=2.5M$ at $(a=1,\alpha=0.2)$. Throughout, $M=1$.}
\label{tab:1}
\\
\hline
\rowcolor{orange!50}
\diagbox[innerwidth=1.2cm, height=1.2cm, linecolor=black, font=\large\bfseries]{\raisebox{0.1em}{$\alpha$}}{\raisebox{-0.6em}{$a$}} 
& \textbf{0} & \textbf{0.2} & \textbf{0.4} & \textbf{0.6} & \textbf{0.8} & \textbf{1.0} \\
\hline
\endfirsthead

\hline
\rowcolor{orange!50}
\diagbox[innerwidth=1.2cm, height=1.2cm, linecolor=black, font=\large\bfseries]{\raisebox{0.1em}{$\alpha$}}{\raisebox{-0.6em}{$a$}}
& \textbf{0} & \textbf{0.2} & \textbf{0.4} & \textbf{0.6} & \textbf{0.8} & \textbf{1.0} \\
\hline
\endhead

\textbf{0}     & 1.00000 & 1.20000 & 1.40000 & 1.60000 & 1.80000 & 2.00000 \\
\hline
\textbf{0.05}  & 1.28801 & 1.36489 & 1.52573 & 1.71168 & 1.90650 & 2.10526 \\
\hline
\textbf{0.10}  & 1.46248 & 1.52087 & 1.65996 & 1.83461 & 2.02466 & 2.22222 \\
\hline
\textbf{0.15}  & 1.63212 & 1.68112 & 1.80563 & 1.97091 & 2.15655 & 2.35294 \\
\hline
\textbf{0.20}  & 1.80902 & 1.85208 & 1.96589 & 2.12321 & 2.30475 & 2.50000 \\
\hline

\end{longtable*}

The metric function behavior between horizons—characterized by the depth and location of its minimum—encodes information about the geodesic structure and effective potential for particle motion in these spacetimes. The variations displayed in Fig.~\ref{izzetletelier_metric_function} thus provide not only a catalog of horizon configurations but also geometric intuition for how EM fields and CoS matter compete to sculpt BH causal structures, demonstrating the rich phenomenology accessible in this classical solution to Einstein's equations with exotic matter sources.

\section{Remarkable Physical Features of Letelier BH Immersed in EMU} \label{isec3}

In this section, we investigate the key physical properties of the Letelier BH immersed in EMU. We begin by examining the thermodynamic characteristics, including the Hawking temperature, Bekenstein-Hawking entropy, and heat capacity, all of which are modified by the CoS parameter $\alpha$ and the EMU parameter $a$. Subsequently, we analyze the dynamics of massive test particles, focusing on the effective potential, conserved quantities, and the ISCO radius. The study then extends to scalar perturbations and the computation of QNMs using the sixth-order WKB approximation, which shed light on the stability and ringdown behavior of the BH. Finally, we examine the photon sphere structure, shadow radius, and photon trajectories, elucidating how the interplay between cosmic string matter and the EM background shapes the optical properties of this spacetime. Throughout this analysis, we compare our results with the well-known SBH and RN limits to highlight the novel features introduced by the combined CoS and EMU effects.

\subsection{Thermodynamics} \label{sec3a}

BH thermodynamics has been a cornerstone of theoretical physics since the seminal works of Bekenstein \cite{isz25} and Hawking \cite{isz26}, who established that BHs possess well-defined thermodynamic properties including temperature, entropy, and heat capacity. For modified BH solutions surrounded by exotic matter or immersed in external fields, these thermodynamic quantities acquire corrections that encode information about the surrounding environment \cite{sec3is01,sec3is02}. In this subsection, we derive the thermodynamic properties of the Letelier BH immersed in EMU and analyze how the CoS parameter $\alpha$ and EMU parameter $a$ influence the event horizon, Hawking temperature, entropy, and heat capacity.

Since the spacetime described by Eq.~(\ref{aa4}) is asymptotically bounded rather than asymptotically flat-owing to the global $(1-\alpha)$ factor from the CoS-the surface gravity at the horizon must be defined with care. Following the standard prescription for non-asymptotically flat spacetimes \cite{sec3is03}, the surface gravity is given by
\begin{equation}
    \kappa=-\frac{C}{2} \frac{\partial_r g_{tt}}{\sqrt{-g_{tt}\,g_{rr}}}\Big{|}_{r=r_{+}},\label{ss1}
\end{equation}
where the normalization factor $C$ is determined by
\begin{equation}
    C^2=\lim_{r \to \infty} |g^{tt}|.\label{ss2} 
\end{equation}
For asymptotically flat spacetimes, $C=1$. However, in our case, the metric function approaches $f(r)\to 1-\alpha$ as $r\to\infty$, yielding $C^2=(1-\alpha)^{-1}$ and hence $C=(1-\alpha)^{-1/2}$. Evaluating Eq.~(\ref{ss1}) at the outer horizon $r_+$, we obtain the surface gravity
\begin{align}
    \kappa=\frac{(1-\alpha)^{-1/2}}{2 r_{+}}\left(1-\alpha-\frac{(1-a^2) M^2}{r^2_{+}}\right),\label{ss3}
\end{align}
where the second equality follows from applying the horizon condition~(\ref{condition}).

The Hawking temperature, defined as $T=\kappa/(2\pi)$, takes the form
\begin{equation}
    T=\frac{\kappa}{2\pi}=\frac{(1-\alpha)^{-1/2}}{4 \pi r_{+}}\left(1-\alpha-\frac{(1-a^2) M^2}{r^2_{+}}\right).\label{ss4}
\end{equation}
This expression reduces to several known limits. In the SBH limit ($a=1$), the horizon radius simplifies to $r_{+}=2M/(1-\alpha)$, and the Hawking temperature becomes
\begin{equation}
T_{\rm LBH}=\frac{(1-\alpha)^{3/2}}{8 \pi M},\label{ss4b}
\end{equation}
which is the Letelier BH temperature. Setting $\alpha=0$ further recovers the standard SBH temperature $T_{\rm SBH}=1/(8\pi M)$. In the opposite limit $\alpha\to 0$ with $a<1$, we obtain the SEBH temperature. The temperature vanishes when the BH becomes extremal, i.e., when the two horizons merge.

\begin{figure}[ht!]
    \centering
    \includegraphics[width=1\linewidth]{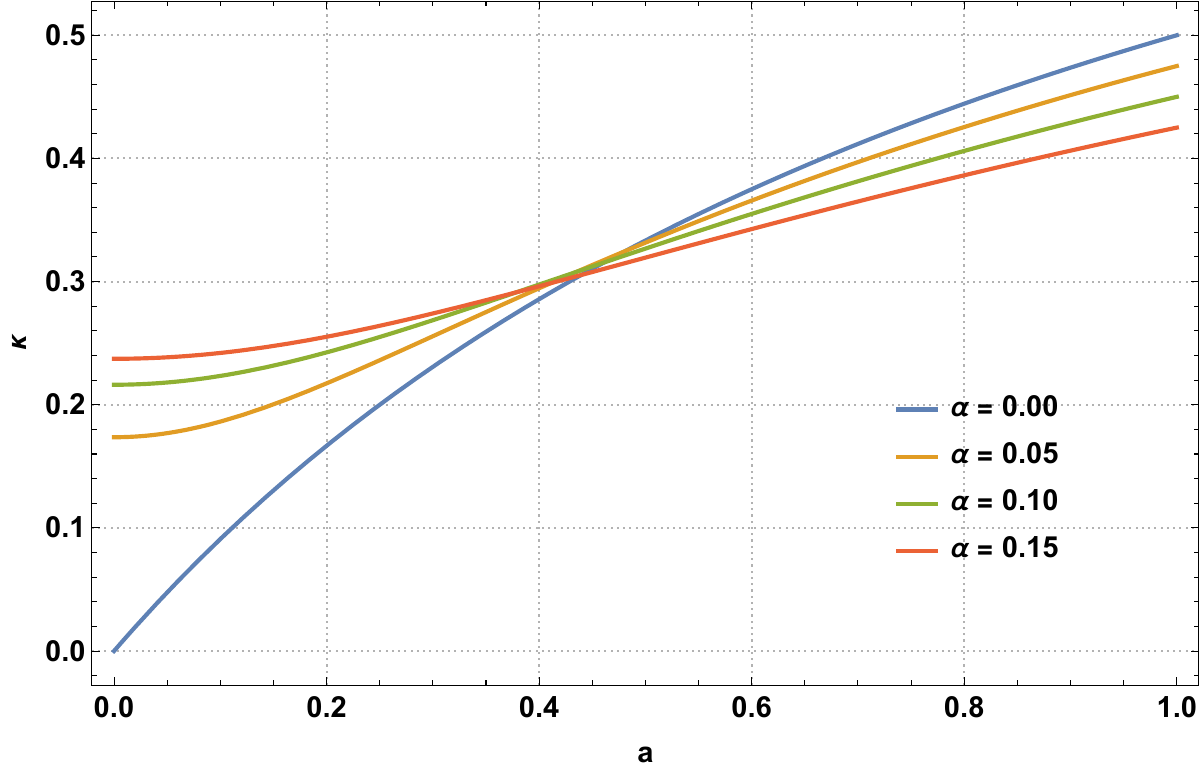}
    \caption{\footnotesize Surface gravity $\kappa$ as a function of EMU parameter $a$ for varying CoS parameter $\alpha$ with $M=1$. The surface gravity increases with $a$ (approaching the SBH limit) and decreases with $\alpha$ (stronger CoS effect). All curves show smooth monotonic behavior across the parameter space.}
    \label{fig:surface-gravity}
\end{figure}

Figure~\ref{fig:surface-gravity} displays the surface gravity $\kappa$ as a function of the EMU parameter $a$ for various values of the CoS parameter $\alpha$. Several features are evident from this plot. First, for fixed $\alpha$, the surface gravity increases monotonically with $a$, reaching its maximum at the SBH limit ($a=1$). This behavior reflects the fact that smaller $a$ values (stronger EM backgrounds) produce larger horizon radii through the $(1-a^2)M^2/r^2$ term, which in turn reduces $\kappa$. Second, for fixed $a$, increasing $\alpha$ decreases the surface gravity, as the CoS dilutes the effective gravitational field strength. Third, all curves exhibit smooth, monotonic behavior without any critical points, indicating that the surface gravity remains well-defined across the entire parameter space where horizons exist.

The horizon area is computed from the induced metric on the $t=\text{const}$, $r=r_+$ surface:
\begin{equation}
    \mathcal{A}(r_+)=\lim_{r \to r_+}\int\int \sqrt{g_{\theta\theta}\,g_{\phi\phi}}\, d\theta\, d\phi=4\pi r^2_{+}.\label{ss5}
\end{equation}
The Bekenstein-Hawking entropy then follows from the area law \cite{isz25}:
\begin{equation}
    S=\frac{\mathcal{A}}{4}=\pi r^2_{+},\label{ss6}
\end{equation}
which retains the standard form but is affected by both $\alpha$ and $a$ through their influence on the horizon radius $r_+$ given in Eq.~(\ref{aa7}). Since $r_+$ increases with both $\alpha$ and $a$, the entropy correspondingly increases as either parameter grows.

The heat capacity at constant charge (or in this case, constant $\alpha$ and $a$) provides crucial information about thermodynamic stability and phase transitions \cite{sec3is04,sec3is05}. It is defined as
\begin{equation}
    C_{\rm heat}=T\left(\frac{dS}{dT}\right).\label{ss7a}
\end{equation}
Evaluating this expression using Eqs.~(\ref{ss4}) and (\ref{ss6}), we obtain
\begin{equation}
    C_{\rm heat}=-2 \pi r_+^2\left(\frac{r_+^2-r^2_c/3}{r_+^2-r^2_c}\right),\label{ss7}
\end{equation}
where we have introduced the critical radius
\begin{equation}
r_c=\sqrt{\frac{3(1-a^2)}{1-\alpha}}\,M.\label{ss7b}
\end{equation}
The heat capacity exhibits a divergence when $r_+=r_c$, signaling a phase transition. For $r_+>r_c$, the heat capacity is negative, indicating thermodynamic instability-the BH loses mass through Hawking radiation and becomes hotter, accelerating its evaporation. For $r_+<r_c$, the heat capacity becomes positive, indicating local thermodynamic stability. This sign change at $r_c$ represents a second-order phase transition analogous to those found in RN and other charged BH systems \cite{sec3is04}.

In the SBH limit ($a=1$), the critical radius vanishes ($r_c=0$), and the heat capacity simplifies to
\begin{equation}
C_{\rm heat}=-2 \pi r_+^2=-\frac{8 \pi M^2}{(1-\alpha)^2},\label{ss7c}
\end{equation}
which is always negative, confirming that the Letelier BH (and by extension, the SBH) is thermodynamically unstable-a well-known result. In the opposite limit $\alpha=0$ (no CoS), the heat capacity becomes
\begin{equation}
C_{\rm heat}=-2 \pi r_+^2\left[\frac{r_+^2- (1-a^2) M^2}{r_+^2-3 (1-a^2) M^2 }\right],\label{ss7d}
\end{equation}
recovering the SEBH result. The presence of both CoS and EMU thus introduces richer thermodynamic behavior than either effect alone.

Using the surface gravity and horizon area, we derive the Smarr-like mass relation
\begin{equation}
    2 T S=\frac{\kappa \mathcal{A}(r_+)}{4 \pi}=M \sqrt{(1-\alpha)^{-1}-(1-a^2)}=\mathcal{M},\label{ss8}
\end{equation}
where $\mathcal{M}$ can be interpreted as the effective thermodynamic mass. This relation generalizes the standard Smarr formula to include contributions from both the CoS and EMU parameters. In the SBH limit ($a=1$, $\alpha=0$), we recover $2TS=M$, the familiar Smarr relation for the SBH.

For completeness, we express the temperature and heat capacity in terms of the entropy $S$:
\begin{align}
    T&=\frac{1}{4 \sqrt{S \pi (1-\alpha)}}\left[1-\alpha-\frac{\pi (1-a^2) M^2}{S}\right],\label{ss9}\\
    C_{\rm heat}&=-2 S \left[\frac{S(1 - \alpha) - \pi (1-a^2) M^2}{S (1 - \alpha)-3 \pi (1-a^2) M^2 }\right].\label{ss10}
\end{align}
These expressions facilitate the study of thermodynamic processes at constant entropy and reveal the explicit dependence of thermodynamic quantities on the BH entropy. The temperature in Eq.~(\ref{ss9}) vanishes when $S=\pi(1-a^2)M^2/(1-\alpha)$, corresponding to the extremal configuration where the two horizons merge. The heat capacity in Eq.~(\ref{ss10}) diverges when $S=3\pi(1-a^2)M^2/(1-\alpha)$, marking the phase transition point.

In summary, the thermodynamics of the Letelier BH immersed in EMU exhibits several distinctive features: (i) the Hawking temperature is suppressed by both increasing CoS density and stronger EM backgrounds; (ii) the entropy follows the area law but is enhanced by both $\alpha$ and $a$ through their effect on the horizon radius; (iii) the heat capacity changes sign at a critical radius $r_c$, indicating a phase transition between thermodynamically stable and unstable regimes; and (iv) the Smarr relation is modified to include contributions from both exotic matter sources. These results demonstrate how CoS and EMU jointly shape the thermodynamic landscape of this BH solution.

\subsection{Dynamics of Test Particles} \label{sec3b}

The motion of test particles around BHs provides valuable information about the spacetime geometry and has direct observational implications for accretion disk physics and gravitational wave astronomy \cite{sec3is06,sec3is07}. In this subsection, we analyze the geodesic motion of massive test particles in the Letelier BH immersed in EMU, deriving the effective potential, conserved quantities, and the ISCO radius.

Since the spacetime described by Eq.~(\ref{aa4}) is static and spherically symmetric, it possesses two Killing vector fields: $\xi_{(t)} \equiv \partial_t$ associated with time-translation invariance, and $\xi_{(\phi)} \equiv \partial_\phi$ associated with axial symmetry. These symmetries give rise to two conserved quantities along particle geodesics. For a test particle of mass $m$, the conserved specific energy is $p_t/m = -\mathcal{E}$, where $\mathcal{E}$ denotes the energy per unit mass, and the conserved specific angular momentum is $p_\phi/m = \mathcal{L}$, where $\mathcal{L}$ is the angular momentum per unit mass.

Due to the spherical symmetry, we may restrict attention to equatorial motion with $\theta = \pi/2$ without loss of generality. The radial equation of motion then takes the form
\begin{equation}
\dot{r}^2 + V_{\rm eff}(r) = \mathcal{E}^2, \label{aa10}
\end{equation}
where $\dot{r} = dr/d\tau$ denotes the derivative with respect to proper time $\tau$, and the effective potential for timelike geodesics is given by
\begin{equation}
    V_{\rm eff}(r)=\left(1+\frac{\mathcal{L}^2}{r^2}\right)f(r).\label{aa11}
\end{equation}
Substituting the metric function from Eq.~(\ref{aa5}), we obtain
\begin{equation}
V_{\rm eff}(r)=\left(1+\frac{\mathcal{L}^2}{r^2}\right)\left(1-\alpha-\frac{2M}{r}+\frac{M^2(1-a^2)}{r^2}\right).\label{aa11b}
\end{equation}
This effective potential reduces to the SBH case when $\alpha=0$ and $a=1$, to the Letelier BH when $a=1$, and to the SEBH when $\alpha=0$.

\begin{figure*}[ht!]
    \centering
    \includegraphics[width=0.48\textwidth]{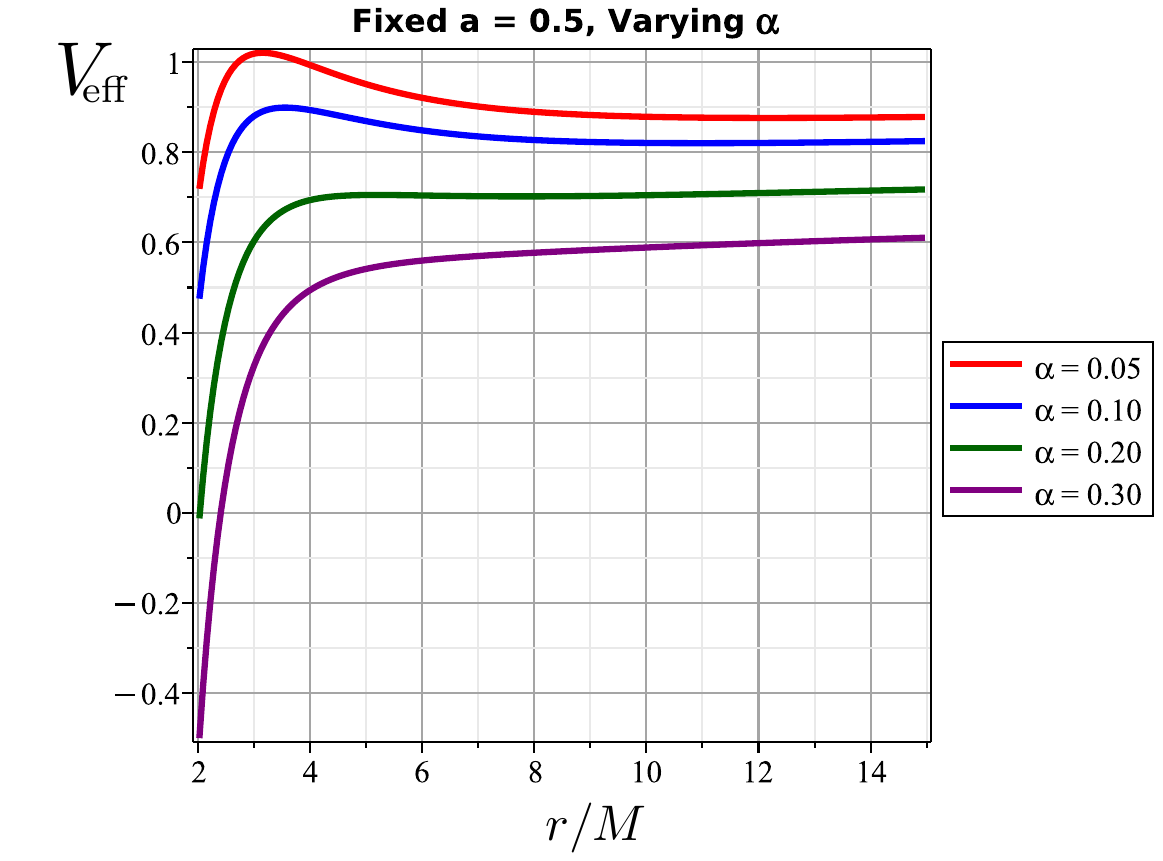}
    \hfill
    \includegraphics[width=0.48\textwidth]{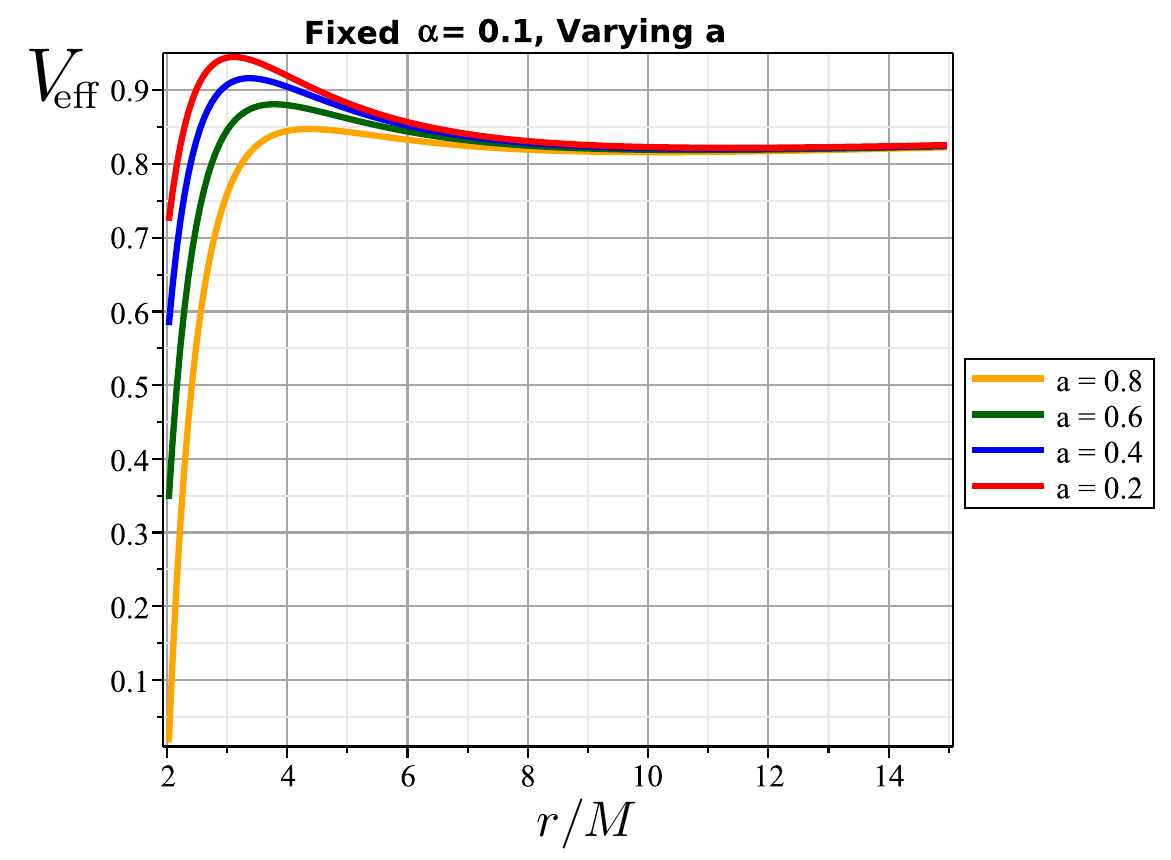}
    \caption{\footnotesize Effective potential $V_{\rm eff}$ for timelike geodesics in the Letelier BH spacetime immersed in EMU with $M=1$ and $\mathcal{L}=4$. \textbf{Left panel:} Fixed EMU parameter $a=0.5$ with varying CoS parameter $\alpha$. The curves correspond to $\alpha=0.05$ (red), $\alpha=0.10$ (blue), $\alpha=0.20$ (green), and $\alpha=0.30$ (purple). As $\alpha$ increases, the asymptotic value $V_{\rm eff}(r\to\infty)\to 1-\alpha$ decreases, and the potential barrier is progressively suppressed, indicating weaker gravitational confinement. \textbf{Right panel:} Fixed CoS parameter $\alpha=0.1$ with varying EMU parameter $a$. The curves correspond to $a=0.8$ (yellow), $a=0.6$ (green), $a=0.4$ (blue), and $a=0.2$ (red). All curves asymptotically approach $V_{\rm eff}\to 0.9$, reflecting the universal CoS contribution. As $a$ decreases (stronger EM field), the potential peak is suppressed and shifts inward due to the $(1-a^2)M^2/r^2$ term.}
    \label{fig:time-like-potential}
\end{figure*}

Figure~\ref{fig:time-like-potential} displays the effective potential for massive test particles with specific angular momentum $\mathcal{L}=4M$. The left panel shows the dependence on the CoS parameter $\alpha$ at fixed $a=0.5$. A notable feature is the monotonic decrease in the asymptotic value: since $V_{\rm eff}(r\to\infty) \to 1-\alpha$, increasing $\alpha$ from $0.05$ to $0.30$ lowers the asymptote from $0.95$ to $0.70$. This global reduction reflects the gravitational dilution effect of CoS matter, which weakens the spacetime's binding capacity. The potential barrier height also decreases with increasing $\alpha$, implying that particles require less energy to escape to infinity. For $\alpha=0.30$, the potential becomes negative near the horizon, signaling a qualitatively different regime where gravitational attraction dominates over the centrifugal barrier.

The right panel of Fig.~\ref{fig:time-like-potential} reveals the complementary effect of the EMU parameter $a$ at fixed $\alpha=0.1$. All curves share the same asymptotic value $V_{\rm eff}\to 0.9$, confirming that the CoS parameter alone governs the far-field behavior. However, the near-horizon structure varies significantly with $a$: as $a$ decreases from $0.8$ to $0.2$ (stronger EM fields), the potential peak is suppressed and shifts inward. This behavior arises from the attractive $(1-a^2)M^2/r^2$ term in the metric function~(\ref{aa5}), which enhances gravitational confinement for $a<1$. The case $a=0.2$ exhibits the lowest and innermost peak, indicating that strong EM backgrounds create tighter potential wells capable of trapping particles more effectively near the BH.

Circular orbits occur at extrema of the effective potential, where $V'_{\rm eff}(r)=0$. Among these, the ISCO represents the smallest radius at which a particle can maintain stable circular motion around the BH. The ISCO marks the inner edge of geometrically thin accretion disks and plays a central role in determining the radiative efficiency of accreting BHs \cite{isz32,isz33}. The conditions for circular orbits are
\begin{equation}
\mathcal{E}^2=V_{\rm eff}(r),\qquad V'_{\rm eff}(r)=0,\label{aa11c}
\end{equation}
while the ISCO additionally requires $V''_{\rm eff}(r_{\rm ISCO}) = 0$. From the first two conditions, we obtain the specific angular momentum and energy for circular orbits:
\begin{align}
    \mathcal{L}^2&=r^2\frac{\frac{M}{r}-\frac{M^2}{r^2}(1-a^2)}{1-\alpha-\frac{3 M}{r}+\frac{2 M^2}{r^2}(1-a^2)},\label{aa12}\\
    \mathcal{E}^2&=\frac{\left(1-\alpha-\frac{2 M}{r}+\frac{2 M^2}{r^2}(1-a^2)\right)^2}{1-\alpha-\frac{3 M}{r}+\frac{2 M^2}{r^2}(1-a^2)}.\label{aa13}
\end{align}
These expressions generalize the well-known SBH results and reduce to them when $\alpha=0$ and $a=1$.

The ISCO radius is determined by the cubic equation
\begin{equation}
    (1-\alpha)\,r^3-6 M r^2+9\, (1-a^2) M^2 r-4 (1-a^2)^2 M^3=0.\label{aa14}
\end{equation}
In the SBH limit ($\alpha=0$, $a=1$), this equation reduces to $r^3-6Mr^2=0$, yielding the familiar result $r_{\rm ISCO}=6M$. For general $\alpha$ and $a$, Eq.~(\ref{aa14}) must be solved numerically.

\begin{figure}[ht!]
    \centering
    \includegraphics[width=1\linewidth]{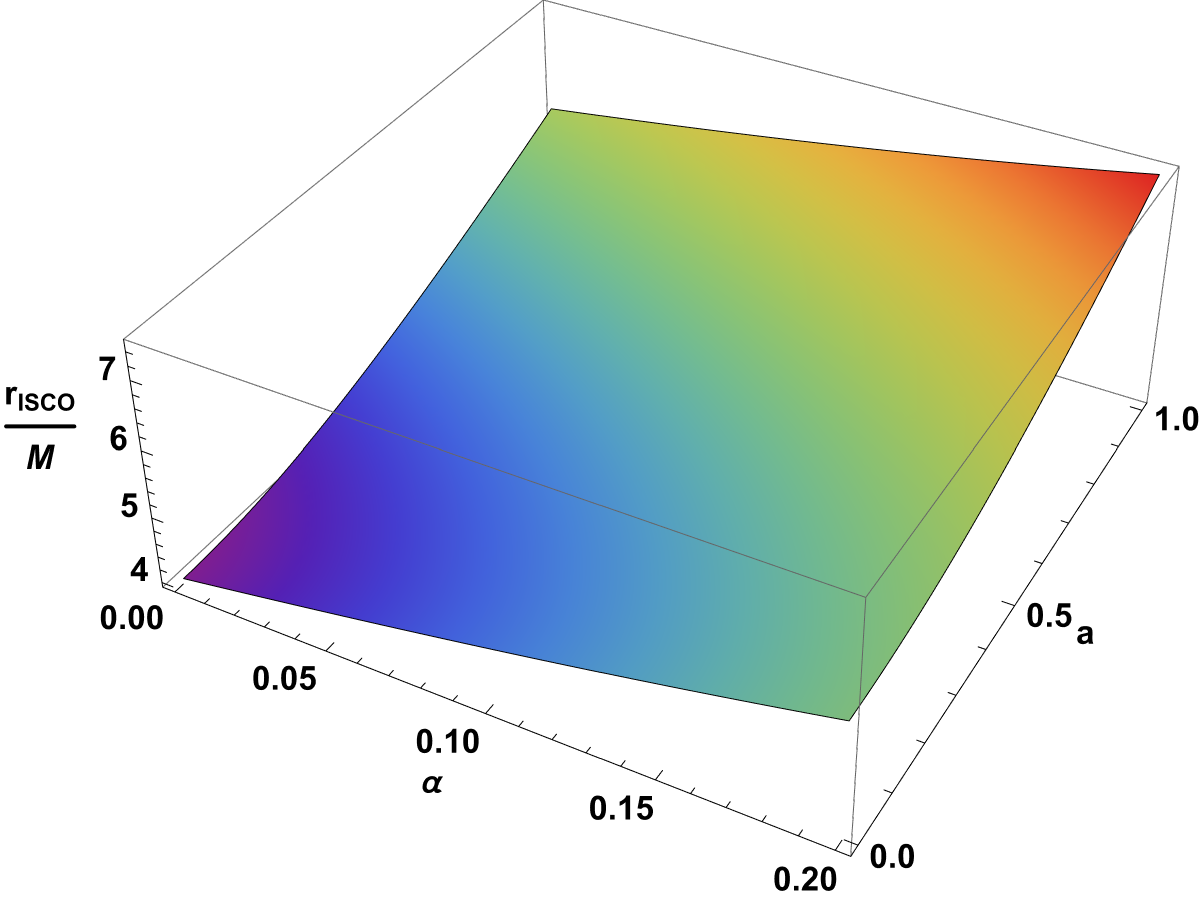}
    \caption{\footnotesize Three-dimensional plot of the ISCO radius as a function of $(\alpha,a)$ with $M=1$. The surface shows that $r_{\rm ISCO}$ increases with both the CoS parameter $\alpha$ and the EMU parameter $a$. For the SBH limit ($a=1$, $\alpha=0$), we recover $r_{\rm ISCO}=6M$. The CoS effect pushes the ISCO outward through the $(1-\alpha)^{-1}$ scaling, while weaker EM fields ($a\to 1$) also expand the ISCO by eliminating the attractive $(1-a^2)$ contribution.}
    \label{fig:ISCO}
\end{figure}

Figure~\ref{fig:ISCO} presents a three-dimensional plot of the ISCO radius as a function of the parameters $(\alpha, a)$. The surface demonstrates that $r_{\rm ISCO}$ increases monotonically with both parameters. The CoS effect pushes the ISCO outward through the $(1-\alpha)^{-1}$ scaling factor, reflecting the gravitational dilution by cosmic string matter. Similarly, increasing $a$ toward the SBH limit expands the ISCO by eliminating the attractive $(1-a^2)M^2/r^2$ contribution from the EMU. For the extremal RN limit ($a\to 0$), the ISCO approaches the photon sphere, while for large $\alpha$ values approaching unity, the ISCO radius diverges as the effective gravitational strength vanishes.

\subsection{Scalar Perturbations} \label{sec3c}

The stability of BH solutions under external perturbations is a fundamental question in gravitational physics \cite{sec3is08,sec3is09}. Scalar perturbations, governed by the massless Klein-Gordon equation in the BH background, provide the simplest testing ground for stability analysis while capturing essential features of how fields interact with strong gravity. The evolution of such perturbations reveals QNMs—damped oscillations that characterize how BHs return to equilibrium after being disturbed. In this subsection, we derive the effective potential governing scalar perturbations in the Letelier BH immersed in EMU.

The dynamics of a massless scalar field $\Psi$ in the curved spacetime~(\ref{aa4}) is governed by the Klein-Gordon equation
\begin{equation}
    \frac{1}{\sqrt{-g}}\partial_{\mu}\left(\sqrt{-g}\, g^{\mu\nu} \partial_{\nu}\Psi\right)=0.\label{bb1}
\end{equation}
For the metric~(\ref{aa4}), the metric components and determinant are
\begin{align}
g_{\mu\nu}&=\text{diag}\left(-f(r),\,f(r)^{-1},\,r^2,\,r^2 \sin^2 \theta\right),\nonumber\\
g^{\mu\nu}&=\text{diag}\left(-f(r)^{-1},\,f(r),\,r^{-2},\,r^{-2} \sin^{-2} \theta\right),\label{bb1b}
\end{align}
with $g=-r^4 \sin^2 \theta$. Exploiting the spherical symmetry and stationarity of the background, we decompose the scalar field as
\begin{equation}
\Psi(t,r,\theta,\phi)=e^{-i \omega t}\,Y^{\ell}_{m}(\theta, \phi)\frac{\psi(r)}{r},\label{bb1c}
\end{equation}
where $Y^{\ell}_m(\theta,\phi)$ are the spherical harmonics with angular momentum quantum numbers $(\ell,m)$, and $\omega$ is the (generally complex) frequency.

Introducing the tortoise coordinate $r_*$ defined by
\begin{equation}
\frac{dr_*}{dr}=\frac{1}{f(r)},\label{bb1d}
\end{equation}
the radial equation transforms into the Schrödinger-like form
\begin{equation}
    \frac{d^2\psi}{dr_{*}^2}+\left(\omega^2-\mathcal{V}_s\right)\psi(r_{*})=0,\label{bb2}
\end{equation}
where the effective potential for scalar perturbations is
\begin{align}
    \mathcal{V}_s&=\left(\frac{\ell\,(\ell+1)}{r^2}+\frac{2 M}{r^3}-\frac{2 M^2}{r^4}(1-a^2)\right)\nonumber\\
    &\quad\times\left(1-\alpha-\frac{2 M}{r}+\frac{M^2}{r^2}(1-a^2)\right).\label{bb3}
\end{align}
This potential consists of a centrifugal term $\propto \ell(\ell+1)/r^2$, a mass term $\propto M/r^3$, and an EMU correction $\propto (1-a^2)M^2/r^4$, all modulated by the metric function $f(r)$.

It is convenient to express the potential in dimensionless form. Defining $x=r/M$, we have
\begin{align}
    M^2\mathcal{V}_s&=\left(\frac{\ell\,(\ell+1)}{x^2}+\frac{2}{x^3}-\frac{2(1-a^2)}{x^4}\right)\nonumber\\
    &\quad\times\left(1-\alpha-\frac{2}{x}+\frac{1-a^2}{x^2}\right).\label{bb4}
\end{align}

\begin{figure}[t]
  \centering
  \vspace{0.25cm} %
    \includegraphics[width=1\linewidth]{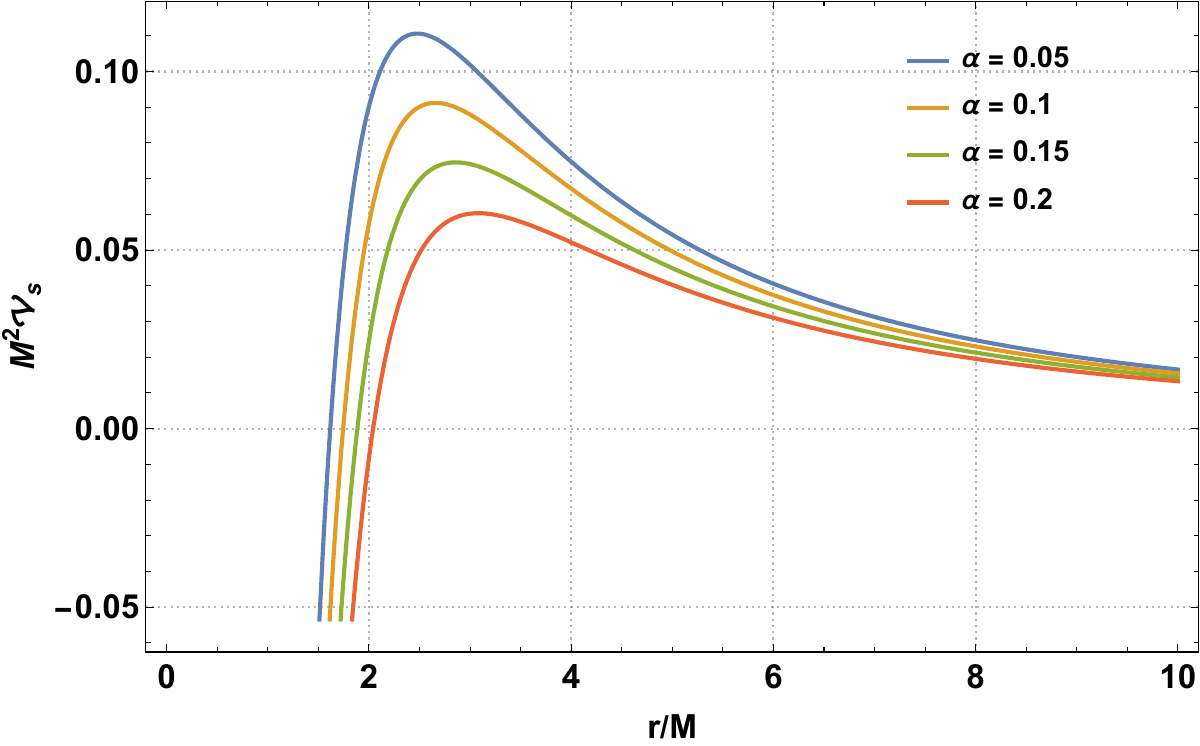}\\
    (i) $a=0.5$\\[2mm]
    \includegraphics[width=1\linewidth]{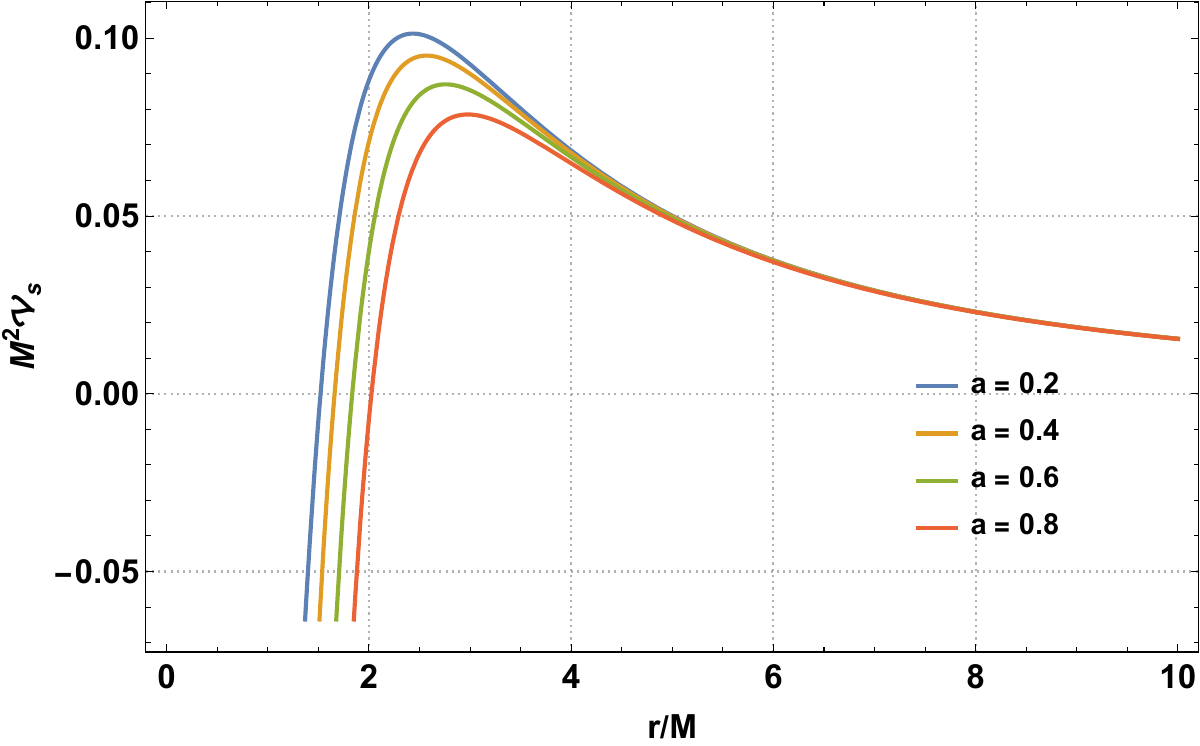}\\
    (ii) $\alpha=0.1$
    \caption{\footnotesize Scalar perturbation potential $M^2\mathcal{V}_s$ as a function of $r/M$ for the $\ell=1$ mode. \textbf{Panel (i):} Fixed $a=0.5$ with varying CoS parameter $\alpha$. The potential peak decreases and shifts outward as $\alpha$ increases. \textbf{Panel (ii):} Fixed $\alpha=0.1$ with varying EMU parameter $a$. The potential peak increases and shifts inward as $a$ decreases (stronger EM field).}
    \label{fig:scalr-potential}
\end{figure}

Figure~\ref{fig:scalr-potential} displays the scalar perturbation potential $M^2\mathcal{V}_s$ for the $\ell=1$ mode. The upper panel shows the variation with $\alpha$ at fixed $a=0.5$. As $\alpha$ increases, the potential peak decreases in height and shifts outward, reflecting the gravitational dilution by CoS matter. The potential also develops a deeper negative region near the horizon for larger $\alpha$, indicating enhanced absorption of scalar waves. The lower panel shows the variation with $a$ at fixed $\alpha=0.1$. As $a$ decreases (stronger EM field), the potential peak increases and moves inward, demonstrating that EM backgrounds enhance the scattering barrier for scalar perturbations.

\subsection{Quasinormal Modes (QNMs)} \label{sec3d}

QNMs are the characteristic oscillation frequencies of perturbed BHs, analogous to the normal modes of vibrating systems but with complex frequencies due to energy dissipation through radiation to infinity and absorption at the horizon \cite{isz36,isz37}. The real part of the QNM frequency $\omega_R$ determines the oscillation rate, while the imaginary part $\omega_I$ (with $\omega_I<0$ for stable modes) determines the decay time. Since QNMs depend solely on the BH parameters, they encode direct information about the spacetime geometry and serve as fingerprints for BH identification in gravitational wave observations \cite{sec3is10}.

We compute the QNM frequencies using the sixth-order WKB approximation, which provides accurate results for the fundamental mode and low overtones when $\ell>n$ \cite{isz38,isz39,isz40}. The WKB formula reads
\begin{equation}
    \frac{i\,(\omega_n-V_0)}{\sqrt{-2V''_0}}-\sum^6_{j=2}\Phi_j=n+\frac{1}{2},\label{bb5}
\end{equation}
where $V_0$ is the maximum of the effective potential, $V''_0$ is its second derivative with respect to the tortoise coordinate evaluated at the maximum, $\Phi_j$ are higher-order WKB corrections, and $n=0,1,2,\ldots$ labels the overtone number. We restrict our analysis to scalar and EM perturbations with $\ell=2$ and $n=0$ (fundamental mode), where the WKB method is most reliable.

Table~\ref{taba3} presents the QNM frequencies for varying EMU parameter $a$ at fixed $\alpha=0.1$. Several trends are evident. First, the real part $\omega_R$ (oscillation frequency) decreases monotonically as $a$ increases from $0.1$ to $0.8$: for scalar perturbations, $\omega_R$ drops from $0.481$ to $0.411$, representing a $15\%$ reduction. This behavior reflects the expansion of the photon sphere with increasing $a$, which lowers the characteristic frequency. Second, the imaginary part $|\omega_I|$ (decay rate) also decreases slightly with increasing $a$, indicating that perturbations decay more slowly in the SBH limit compared to strong EM backgrounds. Third, EM perturbations consistently exhibit lower frequencies and slower decay rates than scalar perturbations, a pattern familiar from SBH and RN studies.

\setlength{\tabcolsep}{10pt}
\renewcommand{\arraystretch}{1.4}
\arrayrulecolor{black}

\begin{longtable*}{|c|c|c|}
\caption{\footnotesize QNM frequencies for scalar and EM perturbations as functions of the EMU parameter $a$ with fixed $\alpha=0.1$, $\ell=2$, $n=0$, and $M=1$. Both the oscillation frequency (real part) and decay rate (imaginary part magnitude) decrease with increasing $a$.}
\label{taba3}
\\
\hline 
\multicolumn{3}{|c|}{$\alpha=0.1$, $n=0$, $\ell=2$, $M=1$} \\ 
\hline
\cellcolor{orange!50}\textbf{$a$} & \cellcolor{orange!50}\textbf{Scalar} & \cellcolor{orange!50}\textbf{EM} \\ 
\hline
\endfirsthead

\hline 
\multicolumn{3}{|c|}{$\alpha=0.1$, $n=0$, $\ell=2$, $M=1$} \\ 
\hline
\cellcolor{orange!50}\textbf{$a$} & \cellcolor{orange!50}\textbf{Scalar} & \cellcolor{orange!50}\textbf{EM} \\ 
\hline
\endhead

$0.1$ & $0.481099-0.0790888i$ & $0.46270-0.0780923i$ \\ 
\hline
$0.2$ & $0.460178-0.0796593i$ & $0.442706-0.0786786i$ \\ 
\hline
$0.3$ & $0.445021-0.0795929i$ & $0.428062-0.0786117i$ \\ 
\hline
$0.4$ & $0.433755-0.0793224i$ & $0.417006-0.078329i$ \\ 
\hline
$0.5$ & $0.425362-0.0790102i$ & $0.40858-0.0779942i$ \\ 
\hline
$0.6$ & $0.419241-0.0787264i$ & $0.402214-0.0776781i$ \\ 
\hline
$0.7$ & $0.415018-0.0785043i$ & $0.397549-0.0774138i$ \\ 
\hline
$0.8$ & $0.411446-0.0783035i$ & $0.394356-0.0772176i$ \\ 
\hline
\end{longtable*}

Table~\ref{taba4} shows the QNM frequencies for varying CoS parameter $\alpha$ at fixed $a=0.4$. The CoS effect is more pronounced than the EMU effect. As $\alpha$ increases from $0.1$ to $0.45$, the real part $\omega_R$ decreases dramatically: for scalar perturbations, it drops from $0.434$ to $0.201$, a reduction exceeding $50\%$. The imaginary part $|\omega_I|$ also decreases substantially, from $0.079$ to $0.029$, indicating that perturbations decay much more slowly in spacetimes with strong CoS. This behavior can be understood from the gravitational dilution effect: larger $\alpha$ weakens the effective gravitational field, expanding the photon sphere and reducing both the oscillation frequency and the decay rate.

\setlength{\tabcolsep}{10pt}
\renewcommand{\arraystretch}{1.4}
\arrayrulecolor{black}

\begin{longtable*}{|c|c|c|}
\caption{\footnotesize QNM frequencies for scalar and EM perturbations as functions of the CoS parameter $\alpha$ with fixed $a=0.4$, $\ell=2$, $n=0$, and $M=1$. Both the oscillation frequency and decay rate decrease substantially with increasing $\alpha$.}
\label{taba4}
\\
\hline 
\multicolumn{3}{|c|}{$a=0.4$, $n=0$, $\ell=2$, $M=1$} \\ 
\hline
\cellcolor{orange!50}\textbf{$\alpha$} & \cellcolor{orange!50}\textbf{Scalar} & \cellcolor{orange!50}\textbf{EM} \\ 
\hline
\endfirsthead

\hline 
\multicolumn{3}{|c|}{$a=0.4$, $n=0$, $\ell=2$, $M=1$} \\ 
\hline
\cellcolor{orange!50}\textbf{$\alpha$} & \cellcolor{orange!50}\textbf{Scalar} & \cellcolor{orange!50}\textbf{EM} \\ 
\hline
\endhead

$0.1$  & $0.433755-0.0793224i$ & $0.417006-0.078329i$ \\ 
\hline
$0.15$ & $0.396529-0.0707102i$ & $0.381806-0.0698584i$ \\ 
\hline
$0.2$  & $0.360611-0.0625944i$ & $0.347787-0.0618718i$ \\ 
\hline
$0.25$ & $0.326019-0.0549753i$ & $0.314966-0.0543697i$ \\ 
\hline
$0.3$  & $0.292777-0.0478532i$ & $0.28336-0.0473525i$ \\ 
\hline
$0.35$ & $0.260909-0.041228i$  & $0.252992-0.0408204i$ \\ 
\hline
$0.4$  & $0.230445-0.0350993i$ & $0.22389-0.0347736i$ \\ 
\hline
$0.45$ & $0.201418-0.0294669i$ & $0.196086-0.029212i$\\ 
\hline

\end{longtable*}

\begin{figure}[ht!]
    \centering
    \includegraphics[width=1\linewidth]{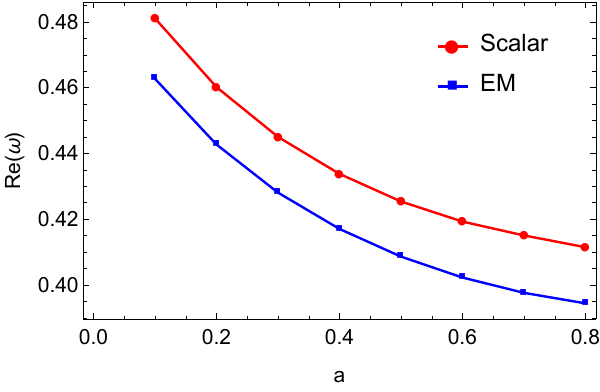}
    \caption{\footnotesize Real part of QNM frequencies (oscillation frequency) as a function of the EMU parameter $a$ for scalar and EM perturbations with $\alpha=0.1$, $\ell=2$, $n=0$, and $M=1$. Both curves decrease monotonically with $a$.}
    \label{fig:placeholder}
\end{figure}

\begin{figure}[ht!]
    \centering
    \includegraphics[width=1\linewidth]{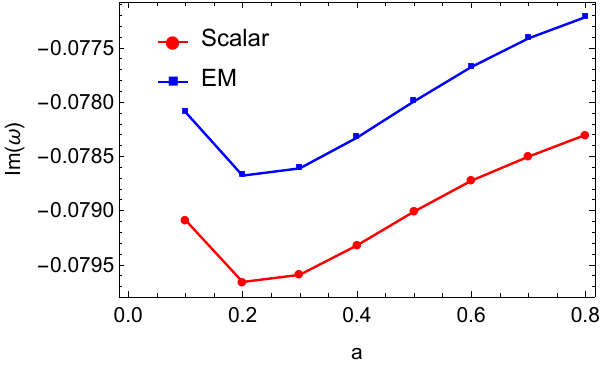}
    \caption{\footnotesize Imaginary part of QNM frequencies (decay rate) as a function of the EMU parameter $a$ for scalar and EM perturbations with $\alpha=0.1$, $\ell=2$, $n=0$, and $M=1$. The decay rate shows weak dependence on $a$.}
    \label{fig:placeholder1}
\end{figure}

\begin{figure}[ht!]
    \centering
    \includegraphics[width=1\linewidth]{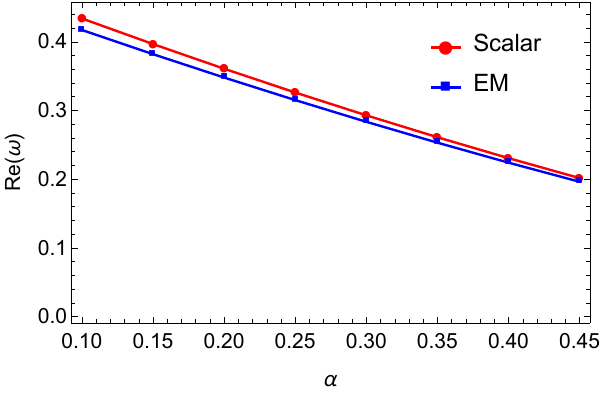}
    \caption{\footnotesize Real part of QNM frequencies as a function of the CoS parameter $\alpha$ for scalar and EM perturbations with $a=0.4$, $\ell=2$, $n=0$, and $M=1$. The oscillation frequency decreases nearly linearly with $\alpha$.}
    \label{fig:placeholder2}
\end{figure}

\begin{figure}[ht!]
    \centering
    \includegraphics[width=1\linewidth]{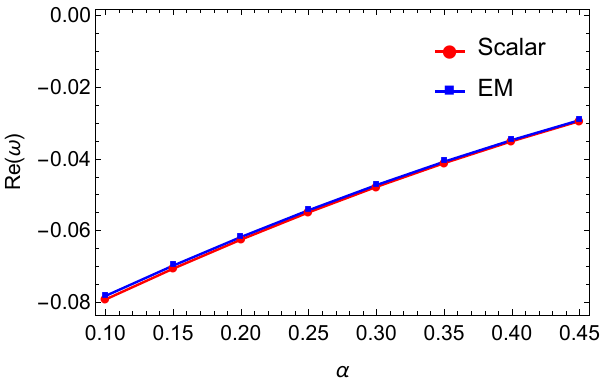}
    \caption{Imaginary part of QNM frequencies as a function of the CoS parameter $\alpha$ for scalar and EM perturbations with $a=0.4$, $\ell=2$, $n=0$, and $M=1$. The decay rate decreases substantially with increasing $\alpha$.}
    \label{fig:placeholder3}
\end{figure}

Figures~\ref{fig:placeholder}--\ref{fig:placeholder3} visualize the QNM frequency dependence on the parameters $a$ and $\alpha$. Figure~\ref{fig:placeholder} shows that the real part $\omega_R$ decreases with $a$ for both scalar and EM perturbations, with scalar modes consistently having higher frequencies. Figure~\ref{fig:placeholder1} displays the imaginary part, which shows a weak decrease with $a$. Figures~\ref{fig:placeholder2} and \ref{fig:placeholder3} demonstrate the much stronger dependence on $\alpha$: both $\omega_R$ and $|\omega_I|$ decrease nearly linearly with $\alpha$, with the slope being steeper for scalar perturbations.

The key findings from the QNM analysis are:
\begin{itemize}
    \item The imaginary part remains negative for all parameter values studied, confirming the stability of the Letelier BH immersed in EMU under scalar and EM perturbations.
    \item Increasing either the EMU parameter $a$ or the CoS parameter $\alpha$ reduces the oscillation frequency $\omega_R$, reflecting the expansion of the photon sphere.
    \item The CoS parameter $\alpha$ has a stronger effect on QNM frequencies than the EMU parameter $a$, with $\alpha$ variations producing changes exceeding $50\%$ compared to $\sim 15\%$ for $a$ variations.
    \item The decay rate $|\omega_I|$ decreases with both parameters, indicating longer-lived ringdown signals for BHs with strong CoS or weak EM backgrounds.
\end{itemize}

\subsection{Photon Sphere, shadow and trajectories} \label{sec3e}

The photon sphere represents a critical surface in BH spacetimes where gravity is strong enough to bend light into circular orbits \cite{sec3is11,sec3is12}. Although these orbits are unstable—small perturbations cause photons to either fall into the BH or escape to infinity—the photon sphere determines the boundary of the BH shadow and plays a central role in gravitational lensing phenomena. In this subsection, we derive the photon sphere radius for the Letelier BH immersed in EMU, analyze the effective potential and radial force for null geodesics, and study the orbital equation governing photon trajectories.

For massless particles, the radial equation of motion takes the form $\dot{r}^2=\mathrm{E}^2-V_{\rm eff}$, where the effective potential is
\begin{equation}
    V_\text{eff}=\frac{\mathrm{L}^2}{r^2}\,f(r).\label{dd0}
\end{equation}
Unlike the timelike case~(\ref{aa11}), the null geodesic potential lacks the constant term proportional to unity, reflecting the massless nature of photons. Substituting the metric function~(\ref{aa5}), we obtain
\begin{equation}
V_{\rm eff}=\frac{\mathrm{L}^2}{r^2}\left(1-\alpha-\frac{2M}{r}+\frac{M^2(1-a^2)}{r^2}\right).\label{dd0b}
\end{equation}

\begin{figure*}[ht!]
    \centering
    \includegraphics[width=0.45\textwidth]{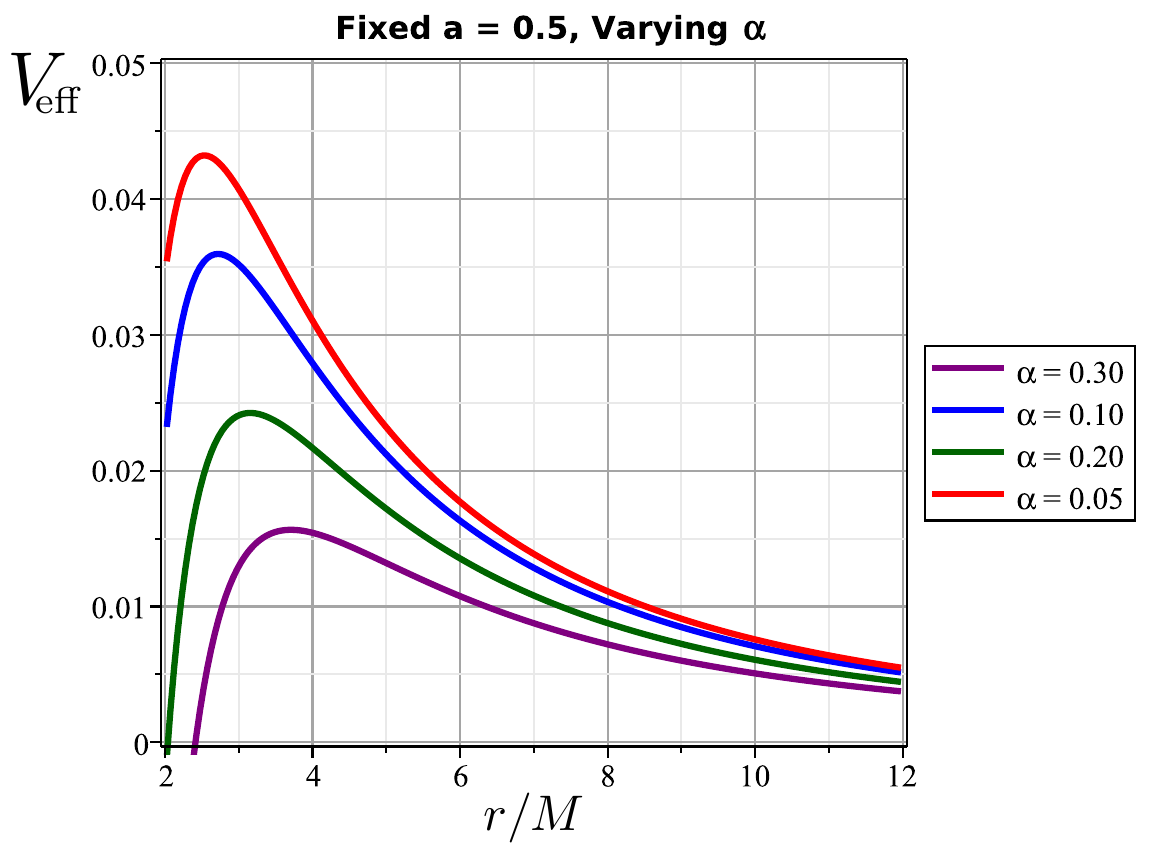}    \includegraphics[width=0.45\textwidth]{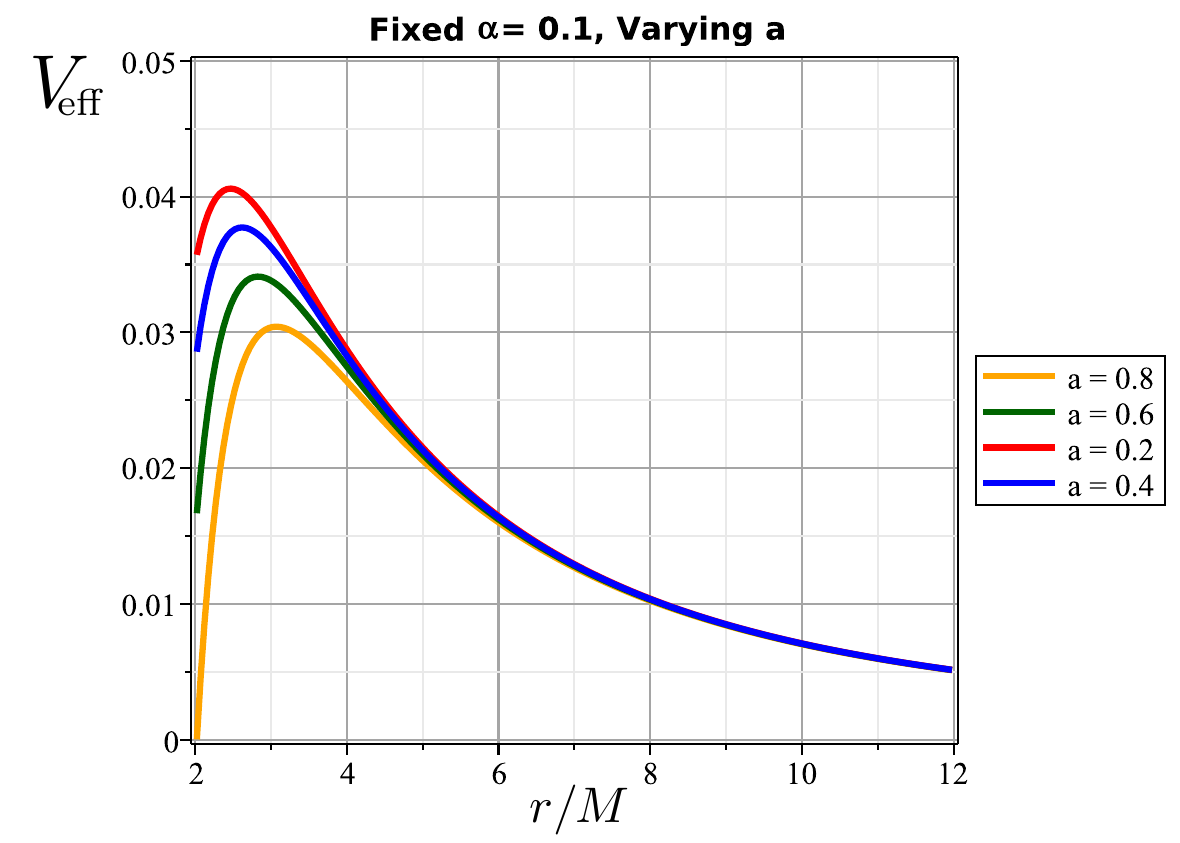}
    \caption{\footnotesize Effective potential $V_{\rm eff}$ for null geodesics in the Letelier BH spacetime immersed in EMU with $M=1$ and $\mathrm{L}=1$. The peak of each curve marks the photon sphere radius $r_{\rm ph}$. \textbf{Left panel:} Fixed $a=0.5$ with varying CoS parameter $\alpha=0.05$ (red), $0.10$ (blue), $0.20$ (green), and $0.30$ (purple). As $\alpha$ increases, the peak decreases and shifts outward, indicating gravitational dilution by CoS matter. \textbf{Right panel:} Fixed $\alpha=0.1$ with varying EMU parameter $a=0.8$ (yellow), $0.6$ (green), $0.4$ (blue), and $0.2$ (red). As $a$ decreases, the peak increases and shifts inward due to enhanced EM confinement. All curves approach $V_{\rm eff}\to 0$ as $r\to\infty$.}
    \label{fig:null-potential}
\end{figure*}

Figure~\ref{fig:null-potential} displays the null geodesic effective potential for photons with angular momentum $\mathrm{L}=M$. A key distinction from the timelike case is that $V_{\rm eff}\to 0$ as $r\to\infty$, characteristic of massless particles. The peak of each curve corresponds to the unstable photon sphere radius $r_{\rm ph}$, where circular photon orbits exist but are unstable against radial perturbations.

The left panel demonstrates the CoS effect at fixed $a=0.5$. As $\alpha$ increases from $0.05$ to $0.30$, two trends emerge: the peak height decreases from approximately $0.043$ to $0.016$, and the peak position shifts outward. This behavior follows directly from the photon sphere formula~(\ref{dd2}), where the factor $(1-\alpha)^{-1}$ amplifies $r_{\rm ph}$ as $\alpha$ increases. Physically, CoS matter dilutes the effective gravitational strength, requiring photons to orbit at larger radii where gravitational attraction balances their angular momentum. The reduced peak height indicates that escaping photons encounter a lower potential barrier, making photon capture less efficient for BHs embedded in dense cosmic string environments.

The right panel reveals the complementary EMU effect at fixed $\alpha=0.1$. As $a$ decreases from $0.8$ to $0.2$ (stronger EM backgrounds), the peak height increases and shifts inward. The case $a=0.2$ exhibits the highest peak near $r\approx 2.5M$, while $a=0.8$ shows a lower peak near $r\approx 3.2M$. This trend arises from the $(1-a^2)M^2/r^2$ term in the metric function: smaller $a$ values enhance this attractive contribution, creating tighter photon confinement. The enhanced peak height for small $a$ implies more efficient photon capture, as incoming light rays must overcome a higher potential barrier to escape.

The photon sphere is determined by the conditions for unstable circular orbits:
\begin{equation}
\mathrm{E}^2=\frac{\mathrm{L}^2}{r^2}\,f(r),\qquad \frac{d}{dr}\left(\frac{f(r)}{r^2}\right)=0. \label{dd1}
\end{equation}
The second condition, equivalent to $2f(r)-rf'(r)=0$, yields the photon sphere radius
\begin{equation}
    r_{\rm ph}=\frac{3 M}{2 (1-\alpha)}\left[1+\sqrt{1-\frac{8 (1-\alpha)(1-a^2)}{9}}\right].\label{dd2}
\end{equation}
In the SBH limit ($\alpha=0$, $a=1$), this reduces to $r_{\rm ph}=3M$, as expected. The CoS parameter enters through the $(1-\alpha)^{-1}$ prefactor, which expands the photon sphere as $\alpha$ increases. The EMU parameter appears in the square root term; as $a$ decreases, the argument under the square root decreases, reducing $r_{\rm ph}$. For the extremal case $a=0$ and $\alpha=0$, corresponding to pure BR geometry, we obtain $r_{\rm ph}=2M$.

The BH shadow, as observed by a static observer at radial position $r_O$, has an apparent radius given by \cite{isz34}
\begin{equation}
    R_{\rm sh}=r_{\rm ph}\sqrt{\frac{f(r_{O})}{f(r_{\rm ph})}}=\sqrt{\frac{1-\alpha-\frac{2M}{r_{O}}+\frac{M^2}{r^2_O}(1-a^2)}{1-\alpha-\frac{2M}{r_{\rm ph}}+\frac{M^2}{r^2_{\rm ph}}(1-a^2)}}\,r_{\rm ph}.\label{shadow}
\end{equation}
For a static observer at infinity ($r_O\to\infty$), this radius simplifies to
\begin{equation}
R_{\rm sh}=r_{\rm ph}\frac{\sqrt{1-\alpha}}{\sqrt{1-\alpha-\frac{2M}{r_{\rm ph}}+\frac{M^2}{r^2_{\rm ph}}(1-a^2)}}.    
\end{equation}
The shadow radius encodes information about both the CoS and EMU parameters, offering a potential observational probe for these effects through EHT-type measurements \cite{isz09,isz10}.

The radial force experienced by photons in the gravitational field provides additional physical insight into photon dynamics. Defining
\begin{equation}
    \mathrm{F}_\text{ph}=-\frac{1}{2}\,\partial_r V_\text{eff},\label{dd3}
\end{equation}
and substituting Eq.~(\ref{dd0}), we obtain
\begin{equation}
    \mathrm{F}_\text{ph}=\frac{\mathrm{L}^2}{r^3}\,\left[1-\alpha-\frac{3M}{r}+\frac{2M^2(1-a^2)}{r^2}\right].\label{dd4}
\end{equation}
The force vanishes at the photon sphere, where gravitational attraction exactly balances the centrifugal repulsion. For $r>r_{\rm ph}$, the force is positive (repulsive), while for $r<r_{\rm ph}$, the force becomes negative (attractive), consistent with the unstable nature of photon sphere orbits.

Several limiting cases are instructive. In the absence of EMU ($a=1$), the radial force simplifies to
\begin{equation}
\mathrm{F}_\text{ph}=\frac{\mathrm{L}^2}{r^3}\left(1-\alpha-\frac{3M}{r}\right),\label{dd5}
\end{equation}
which describes photon dynamics in the pure Letelier BH. In the absence of CoS ($\alpha=0$), we have
\begin{equation}
\mathrm{F}_\text{ph}=\frac{\mathrm{L}^2}{r^3}\left[1-\frac{3M}{r}+\frac{2M^2(1-a^2)}{r^2}\right],\label{dd6}
\end{equation}
corresponding to the SEBH. The combined effects of CoS and EMU shift the zero-crossing of the radial force compared to these limiting cases.

\begin{figure*}[ht!]
\centering
    \includegraphics[width=0.45\linewidth]{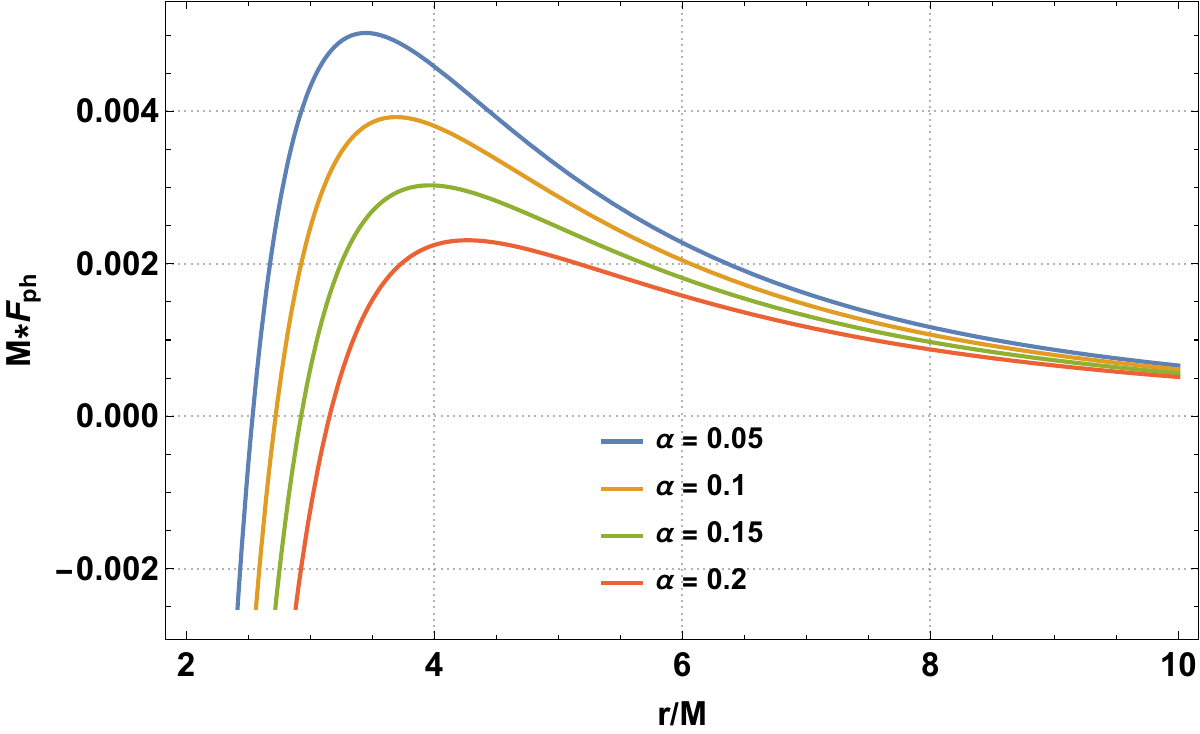}\qquad
    \includegraphics[width=0.45\linewidth]{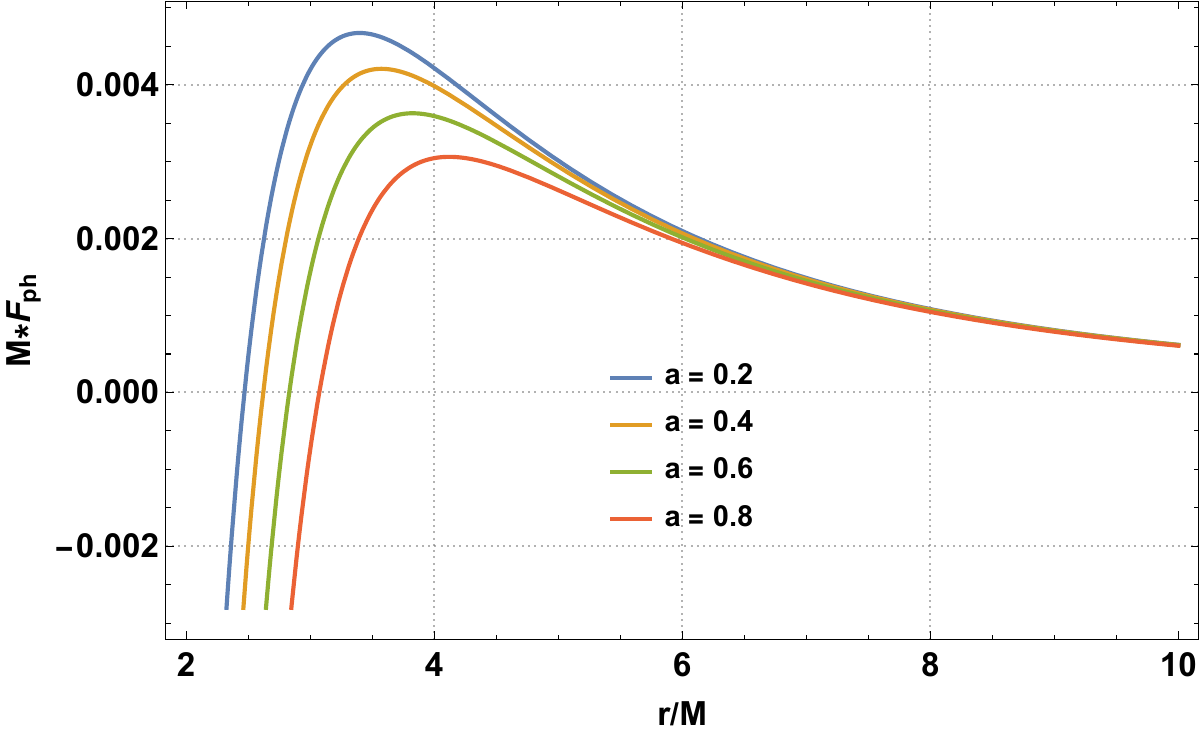}
      \caption{\footnotesize Dimensionless radial force $M\mathrm{F}_{\rm ph}$ for photons as a function of $r/M$ with $\mathrm{L}/M=1$. \textbf{Panel (i):} Fixed $a=0.5$ with varying CoS parameter $\alpha$. The zero-crossing (photon sphere) shifts outward and force magnitude decreases with increasing $\alpha$. \textbf{Panel (ii):} Fixed $\alpha=0.1$ with varying EMU parameter $a$. The zero-crossing shifts inward and near-horizon attraction strengthens as $a$ decreases.}
    \label{fig:force}
\end{figure*}

Figure~\ref{fig:force} displays the dimensionless radial force $M\mathrm{F}_{\rm ph}$ as a function of $r/M$. Panel~(i) shows the variation with $\alpha$ at fixed $a=0.5$: as $\alpha$ increases, the zero-crossing (photon sphere) shifts outward, and the force magnitude decreases at all radii, reflecting the gravitational dilution by CoS matter. Panel~(ii) shows the variation with $a$ at fixed $\alpha=0.1$: as $a$ decreases (stronger EM field), the zero-crossing shifts inward, and the force becomes more strongly attractive near the horizon.

We now turn to the orbital equation governing photon trajectories. From angular momentum conservation, $\dot{\phi}=\mathrm{L}/r^2$, the orbit equation becomes
\begin{equation}
    \left(\frac{dr}{d\phi}\right)^2=\frac{\dot{r}^2}{\dot{\phi}^2}=r^4\left[\frac{1}{\beta^2}-\frac{1}{r^2}\,\left(1-\alpha-\frac{2 M}{r}+\frac{M^2}{r^2}(1-a^2)\right)\right],\label{dd7}
\end{equation}
where $\beta=\mathrm{L}/\mathrm{E}$ is the impact parameter. Introducing the inverse radial variable $u(\phi)=1/r(\phi)$ and differentiating with respect to $\phi$, we obtain the Binet-type equation
\begin{equation}
    \frac{d^2u}{d\phi^2}+(1-\alpha)\,u=3\,M u^2-2 M^2 u^3 (1-a^2).\label{dd9}
\end{equation}
This nonlinear second-order differential equation describes photon trajectories in the combined CoS-EMU geometry. The left-hand side differs from the SBH case through the $(1-\alpha)$ coefficient, while the right-hand side contains both the standard SBH term $3Mu^2$ and the EMU correction $-2M^2u^3(1-a^2)$.

For weak-field trajectories far from the BH, the nonlinear terms on the right-hand side of Eq.~(\ref{dd9}) can be neglected. The resulting linearized equation
\begin{equation}
\frac{d^{2}u}{d\phi^{2}} + (1 - \alpha)\,u = 0\label{dd10}
\end{equation}
has the general solution
\begin{equation}
u(\phi) = \frac{\sin\!\left(\sqrt{1 - \alpha}\,\phi\right)}{\beta},\label{dd11}
\end{equation}
where $\beta$ is the impact parameter. This solution describes oscillatory motion with angular frequency $\sqrt{1-\alpha}$, which differs from unity due to the CoS contribution. The modified frequency implies that photon trajectories in the Letelier BH geometry deviate from straight lines even in the weak-field limit, with the CoS parameter introducing a deficit angle effect analogous to that produced by cosmic strings in flat spacetime \cite{isz25}.


Table~\ref{tab:photon_sphere_results} presents selected photon sphere configurations that illustrate the key physical trends across the parameter space of Letelier BHs immersed in EMU.

\setlength{\tabcolsep}{10pt}
\renewcommand{\arraystretch}{1.6}
\arrayrulecolor{black}

\begin{longtable*}{|c|c|c|c|}
\caption{\footnotesize Selected photon sphere radii for Letelier black holes immersed in EMU. The first three entries show the effect of varying the EM parameter $a$ at nearly vanishing CoS ($\alpha=0.01$), demonstrating monotonic expansion from the extremal RN limit ($a=0$, $r_{\rm ph}=2.040M$) toward the SBH value ($a=1$, $r_{\rm ph}=3M$). The next three entries illustrate the dramatic effect of increasing CoS density $\alpha$ at fixed $a=0.4$, with $r_{\rm ph}$ growing from $2.430M$ at $\alpha=0.05$ to $29.429M$ at $\alpha=0.9$. The final entry provides the SBH benchmark. Throughout, $M=1$.}
\label{tab:photon_sphere_results}
\\
\hline
\rowcolor{orange!50}
\textbf{$a$} & \textbf{$\alpha$} & \textbf{$r_\text{ph}/M$} & \textbf{Physical Configuration} \\
\hline
\endfirsthead

\hline
\rowcolor{orange!50}
\textbf{$a$} & \textbf{$\alpha$} & \textbf{$r_\text{ph}/M$} & \textbf{Physical Configuration} \\
\hline
\endhead

0.0  & 0.01 & 2.040  & Extremal RN with minimal CoS \\
\hline
0.5  & 0.01 & 2.399  & Moderate EM field with minimal CoS \\
\hline
0.95 & 0.01 & 2.964  & Weak EM field (near SBH) with minimal CoS \\
\hline
0.4  & 0.05 & 2.430  & Moderate EM field with weak CoS \\
\hline
0.4  & 0.1  & 2.621  & Moderate EM field with moderate CoS \\
\hline
0.4  & 0.9  & 29.429 & Moderate EM field with strong CoS \\
\hline
1.0  & 0.0  & 3.000  & SBH limit \\
\hline

\end{longtable*}

The entries in Table~\ref{tab:photon_sphere_results} reveal several physical trends. Examining the effect of the EM parameter $a$ at nearly vanishing CoS ($\alpha=0.01$), we observe outward migration of the photon sphere as $a$ increases from 0 to 0.95. At the extremal RN limit ($a=0$), the photon sphere resides at $r_{\rm ph}=2.040M$, slightly larger than the theoretical minimum of $2M$. Weakening the EM field to $a=0.5$ expands the photon sphere to $r_{\rm ph}=2.399M$, a 17.6\% increase. Further reduction to $a=0.95$ pushes the radius to $r_{\rm ph}=2.964M$, approaching the SBH value from below. This monotonic trend confirms that stronger EM fields (smaller $a$) enhance gravitational confinement through the attractive $(1-a^2)M^2/r^2$ metric term.

The entries at fixed $a=0.4$ with varying CoS density demonstrate the gravitational dilution effect of cosmic string matter. At weak CoS ($\alpha=0.05$), the photon sphere sits at $r_{\rm ph}=2.430M$. Increasing to $\alpha=0.1$ expands the radius to $r_{\rm ph}=2.621M$, a modest 7.9\% growth. However, at strong CoS ($\alpha=0.9$), the photon sphere radius jumps to $r_{\rm ph}=29.429M$, nearly an order of magnitude larger than the SBH value. This dramatic expansion occurs because the $(1-\alpha)=0.1$ factor weakens the effective gravitational constant by 90\%, requiring much larger radii to balance centrifugal and gravitational forces. As $\alpha\to 1$, the denominator $2(1-\alpha)$ in Eq.~(\ref{dd2}) approaches zero, causing $r_{\rm ph}\to\infty$ and signaling complete dissolution of gravitational binding. The SBH benchmark ($a=1$, $\alpha=0$) at $r_{\rm ph}=3M$ validates the analytical formula~(\ref{dd2}), with numerical and analytical results agreeing to machine precision.

Table~\ref{tab:photon-sphere} provides a parametric survey of photon sphere locations across the physical parameter space.

\setlength{\tabcolsep}{12pt}
\renewcommand{\arraystretch}{1.6}
\arrayrulecolor{black}

\begin{longtable*}{|>{\columncolor{orange!50}}c|c|c|c|c|c|c|}
\caption{\footnotesize Photon sphere radius $r_{\rm ph}/M$ for Letelier black holes immersed in EMU as functions of the CoS parameter $\alpha$ (rows) and EMU parameter $a$ (columns). The radius increases monotonically with both parameters, ranging from $r_{\rm ph}=2M$ at the extremal RN limit ($a=0$, $\alpha=0$) to $r_{\rm ph}=4M$ at $(a=1,\alpha=0.25)$. Throughout, $M=1$.}
\label{tab:photon-sphere}
\\
\hline
\rowcolor{orange!50}
\diagbox[innerwidth=1.2cm, height=1.2cm, linecolor=black, font=\large\bfseries]{\raisebox{0.1em}{$\alpha$}}{\raisebox{-0.6em}{$a$}} 
& \textbf{0} & \textbf{0.2} & \textbf{0.4} & \textbf{0.6} & \textbf{0.8} & \textbf{1.0} \\
\hline
\endfirsthead

\hline
\rowcolor{orange!50}
\diagbox[innerwidth=1.2cm, height=1.2cm, linecolor=black, font=\large\bfseries]{\raisebox{0.1em}{$\alpha$}}{\raisebox{-0.6em}{$a$}} 
& \textbf{0} & \textbf{0.2} & \textbf{0.4} & \textbf{0.6} & \textbf{0.8} & \textbf{1.0} \\
\hline
\endhead

\textbf{0}     & 2.00000 & 2.07446 & 2.25498 & 2.48489 & 2.73693 & 3.00000 \\
\hline
\textbf{0.05}  & 2.20169 & 2.26599 & 2.43021 & 2.64932 & 2.89621 & 3.15789 \\
\hline
\textbf{0.10}  & 2.41202 & 2.46944 & 2.62119 & 2.83095 & 3.07300 & 3.33333 \\
\hline
\textbf{0.15}  & 2.63720 & 2.68956 & 2.83134 & 3.03290 & 3.27040 & 3.52941 \\
\hline
\textbf{0.20}  & 2.88278 & 2.93123 & 3.06480 & 3.25906 & 3.49229 & 3.75000 \\
\hline
\textbf{0.25}  & 3.15470 & 3.20000 & 3.32665 & 3.51438 & 3.74356 & 4.00000 \\
\hline

\end{longtable*}

The structure of Table~\ref{tab:photon-sphere} reveals several physical patterns. Along any row (fixed $\alpha$, varying $a$), increasing $a$ from 0 to 1 consistently expands the photon sphere. At $\alpha=0.1$, for example, the radius grows from $r_{\rm ph}=2.412M$ at the extremal RN limit to $r_{\rm ph}=3.333M$ at the SBH-with-CoS configuration, representing 38\% expansion. This trend reflects weakening EM confinement as the $(1-a^2)$ term vanishes when $a\to 1$. Down any column (fixed $a$, varying $\alpha$), increasing CoS density produces similar outward migration. At $a=1$ (SBH configuration), adding CoS increases the photon sphere from the classical $r_{\rm ph}=3M$ at $\alpha=0$ to $r_{\rm ph}=3.75M$ at $\alpha=0.2$, a 25\% expansion attributable to gravitational dilution by cosmic string matter.

The diagonal pattern demonstrates cumulative action: configurations with both large $a$ and large $\alpha$ yield the most extended photon spheres, while strong EM fields with minimal CoS produce the tightest confinement. The extremal RN limit ($a=0$, $\alpha=0$) exhibits the smallest photon sphere at exactly $r_{\rm ph}=2M$, precisely two-thirds of the SBH value. The absence of non-monotonic behavior confirms that EM fields and CoS independently weaken gravitational binding without competing or canceling effects. This contrasts with the horizon structure (Table~\ref{izzetletelier_horizons}), where parameter combinations can eliminate horizons or merge them into extremal configurations. All entries correspond to physical photon spheres located outside event horizons, validating the analytical formula~(\ref{dd2}) across the entire parameter space surveyed.

\begin{figure*}[ht!]
    \centering
    \includegraphics[width=0.32\textwidth]{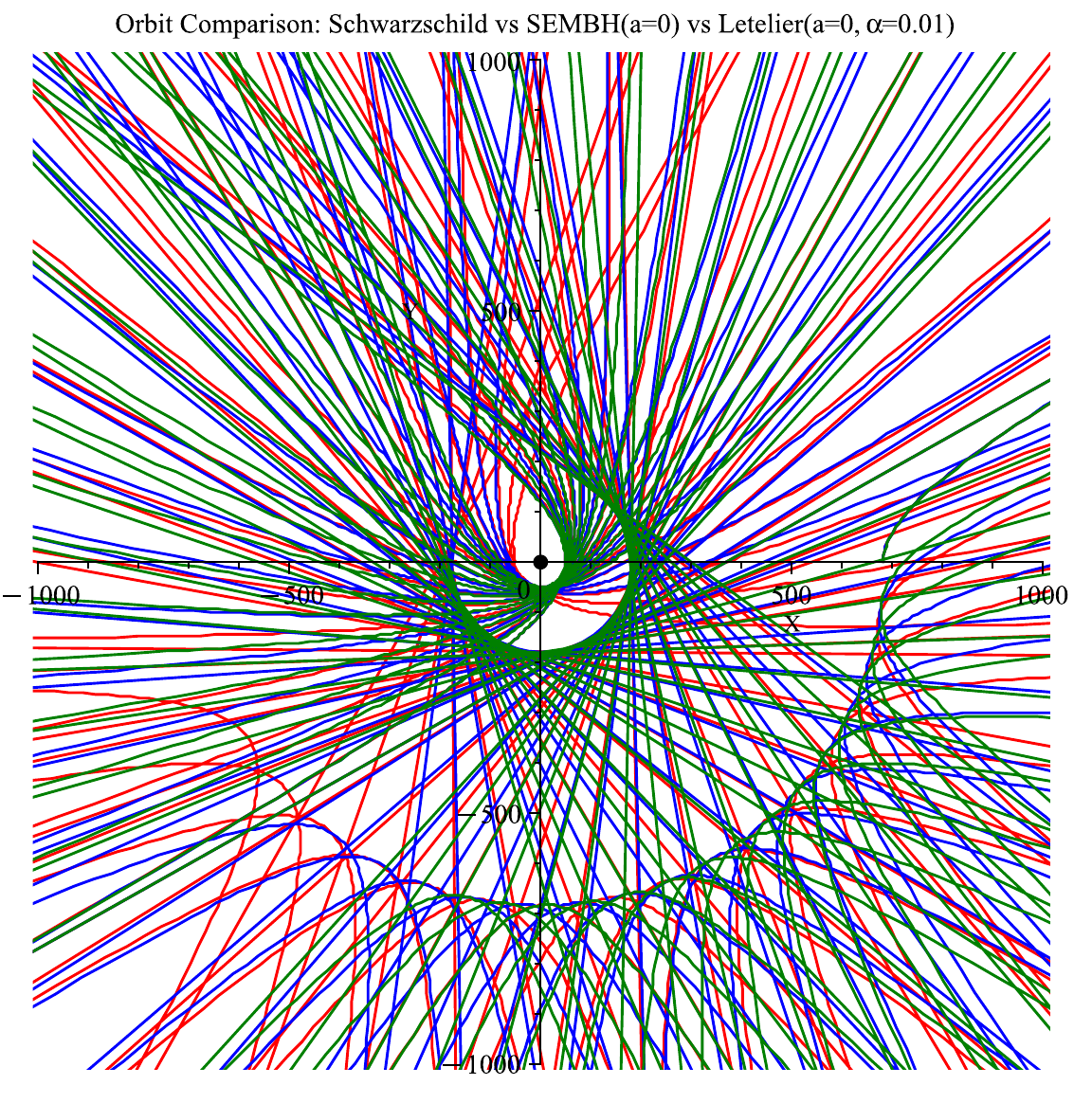}
    \includegraphics[width=0.32\textwidth]{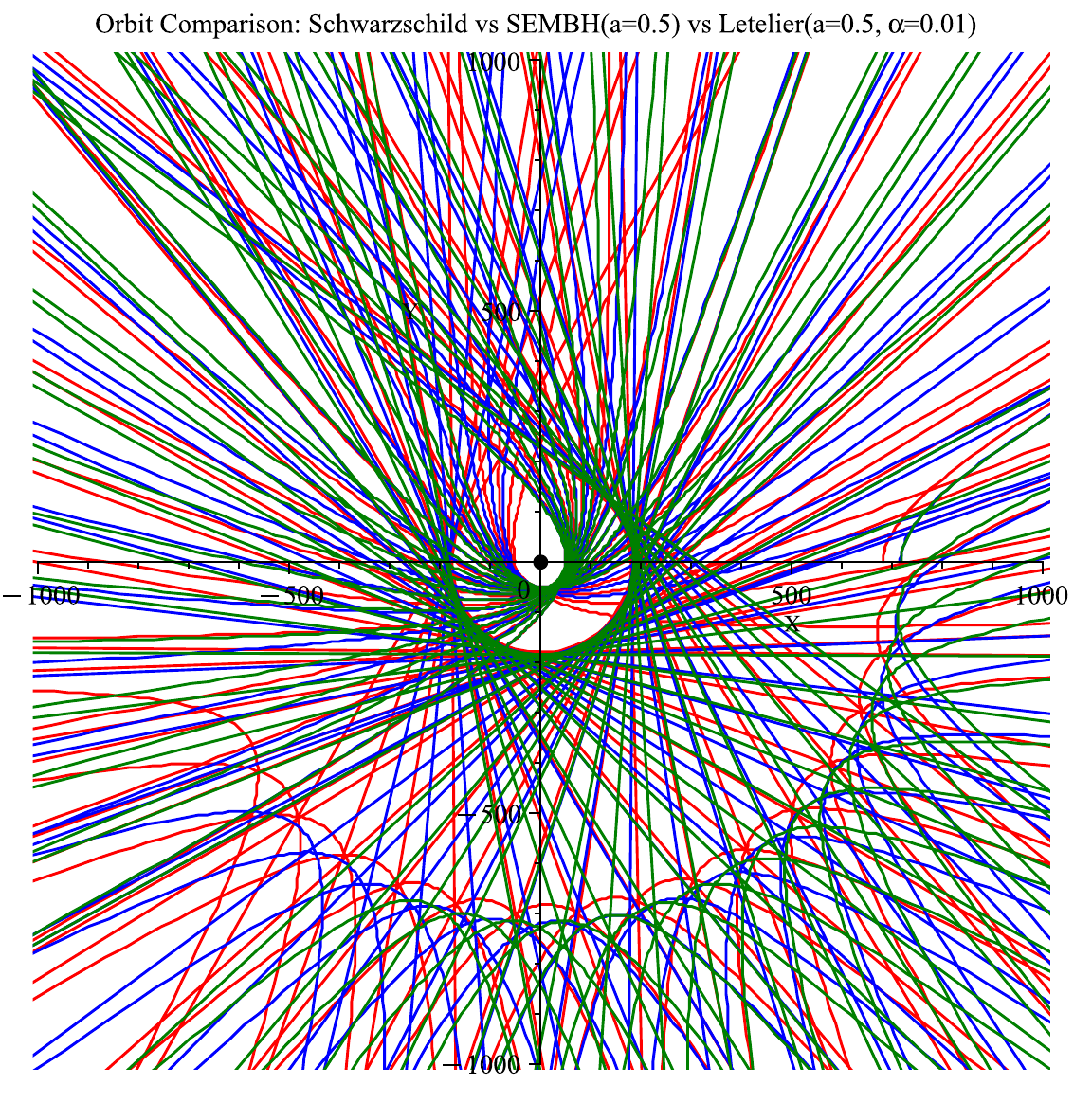}
    \includegraphics[width=0.32\textwidth]{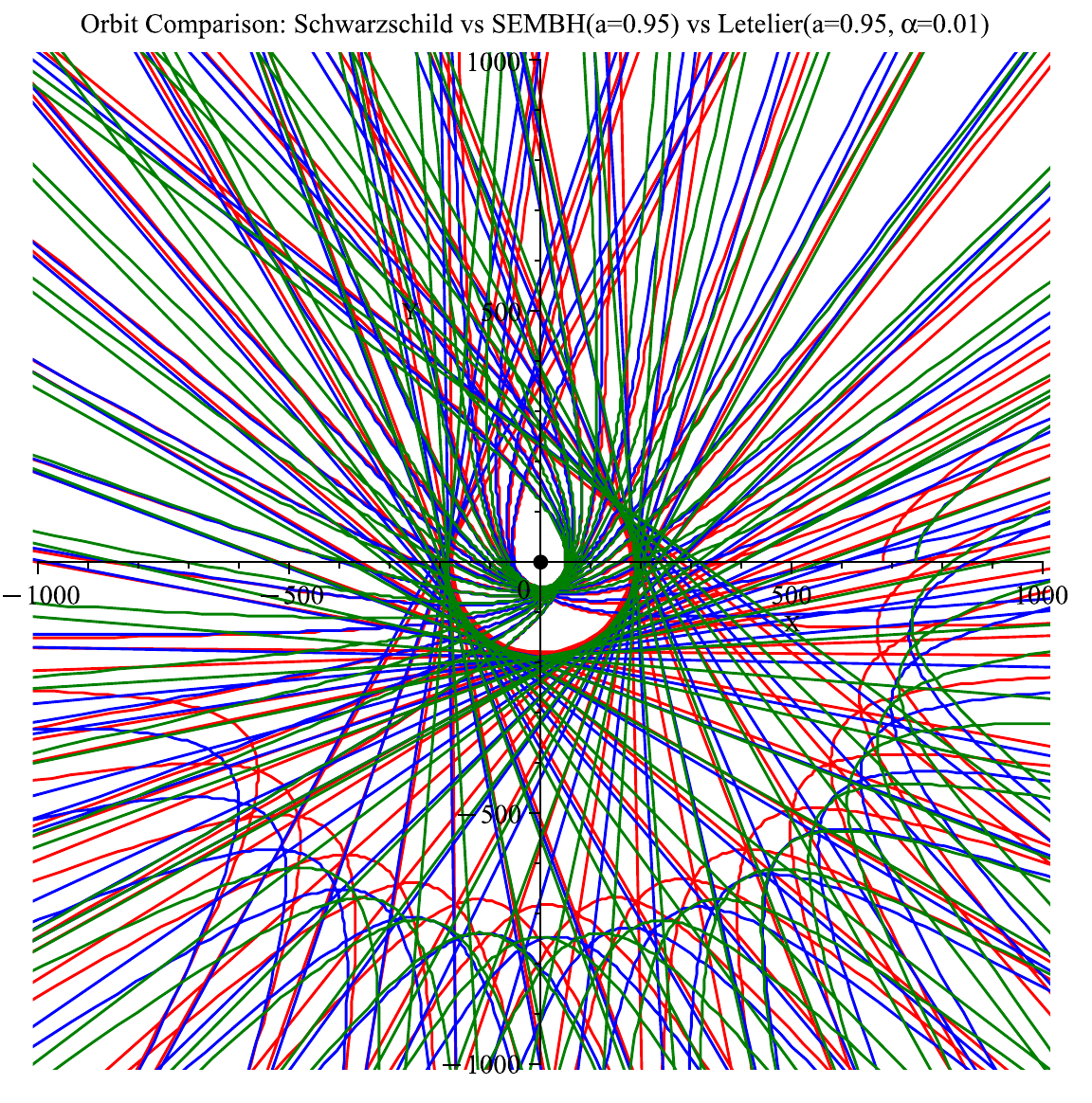}
    \caption{\footnotesize Orbit comparison for fixed CoS parameter $\alpha=0.01$ (nearly vanishing) and varying EMU parameter $a$. Panel (a): $a=0$ (extremal RN limit); Panel (b): $a=0.5$ (moderate EM field); Panel (c): $a=0.95$ (approaching SBH). Red curves: SBH reference ($a=1$, $\alpha=0$); blue curves: SEBH with specified $a$ and $\alpha=0$; green curves: Letelier BH in EMU with specified $a$ and $\alpha=0.01$. As $a$ increases from 0 to 0.95, all orbit types expand outward and approach convergence, reflecting the SBH limit. Panel (a) shows tight configurations with maximum separation, dominated by strong EM confinement. Panel (c) exhibits near-coalescence of all three curves as EM effects become negligible. All panels use $M=1$, $\mathrm{L}=10$.}
    \label{fig:letelier_orbits_varying_a}
\end{figure*}

\begin{figure*}[ht!]
    \centering
    \includegraphics[width=0.32\textwidth]{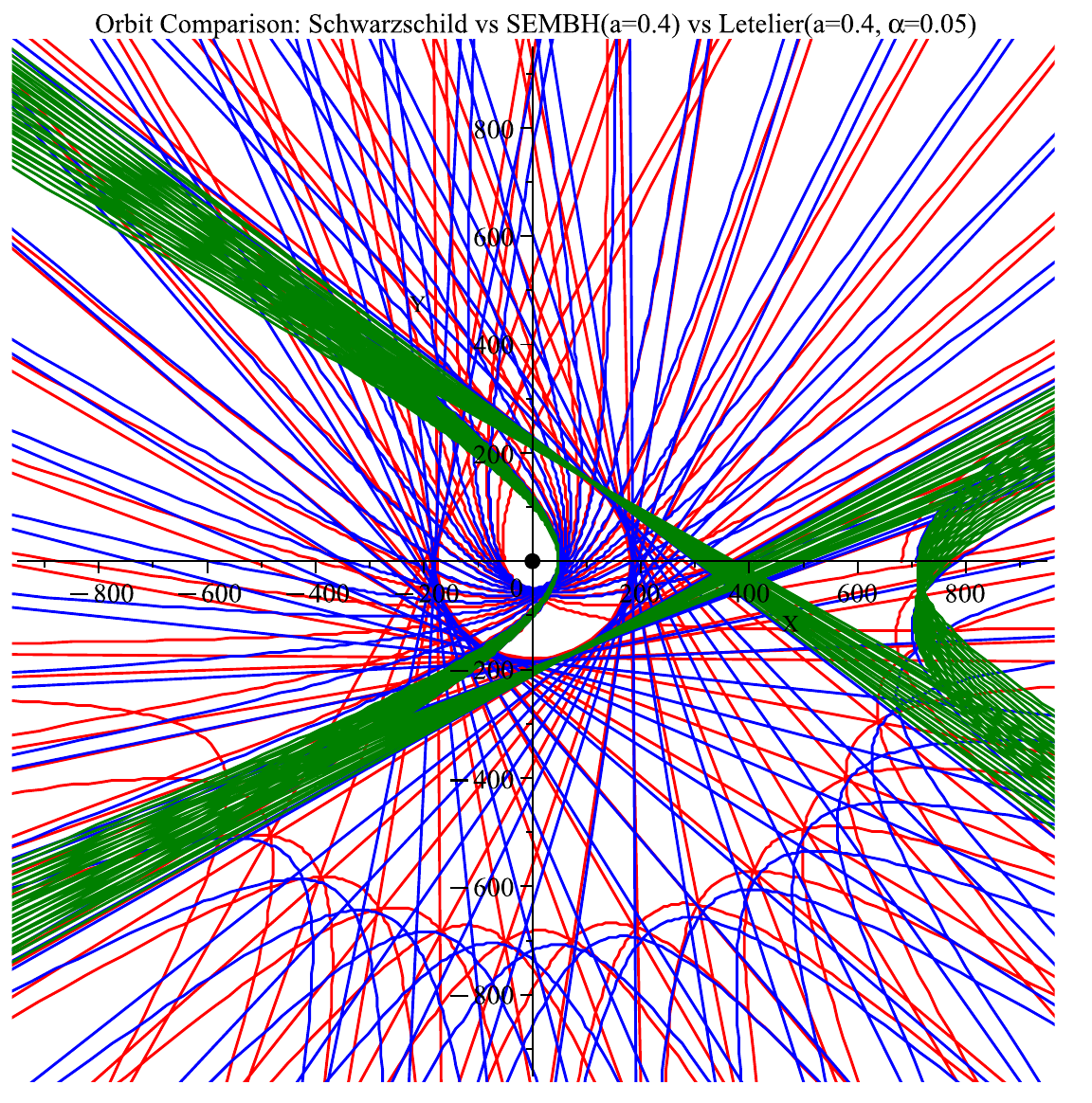}
    \includegraphics[width=0.32\textwidth]{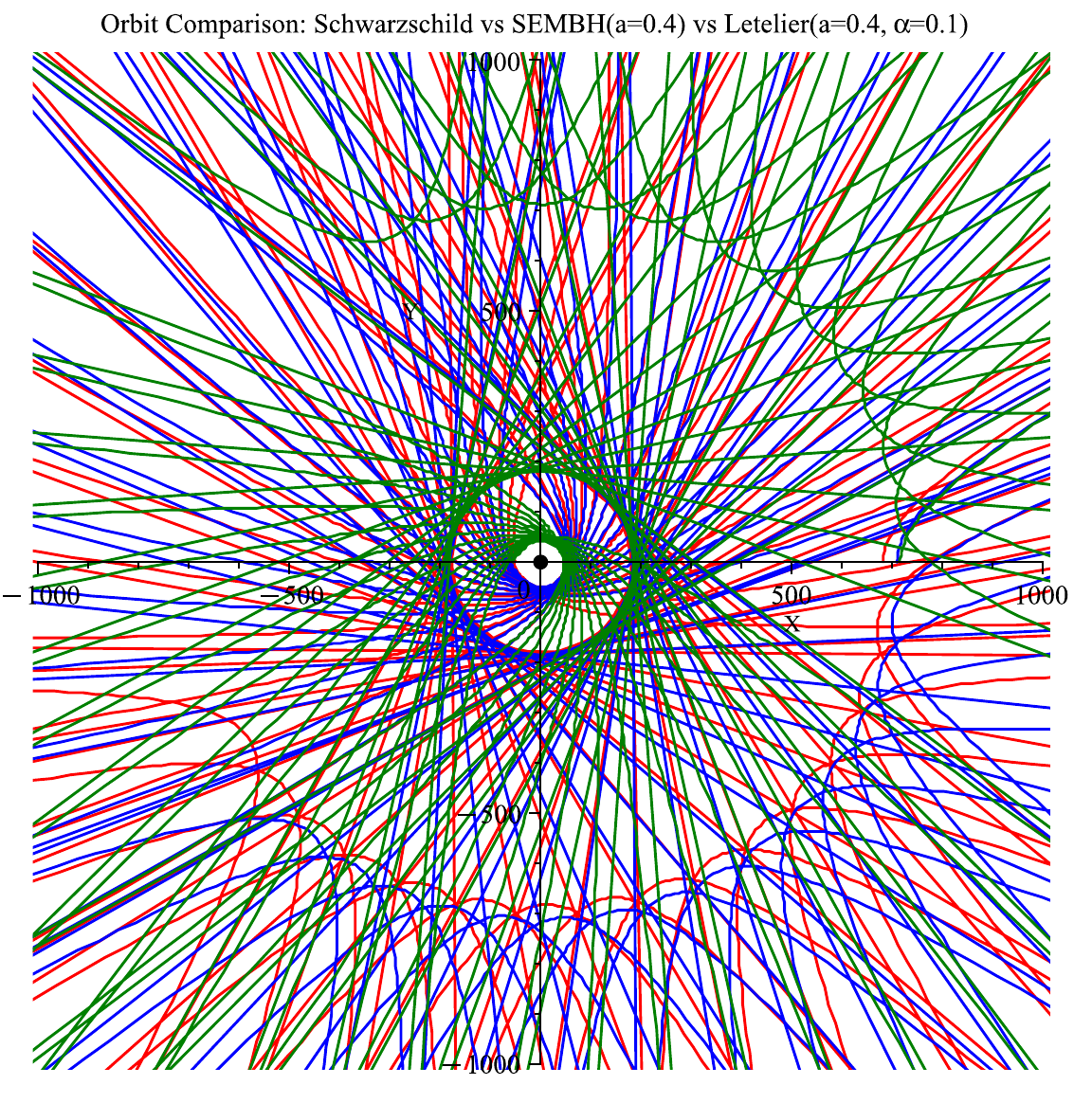}
    \includegraphics[width=0.32\textwidth]{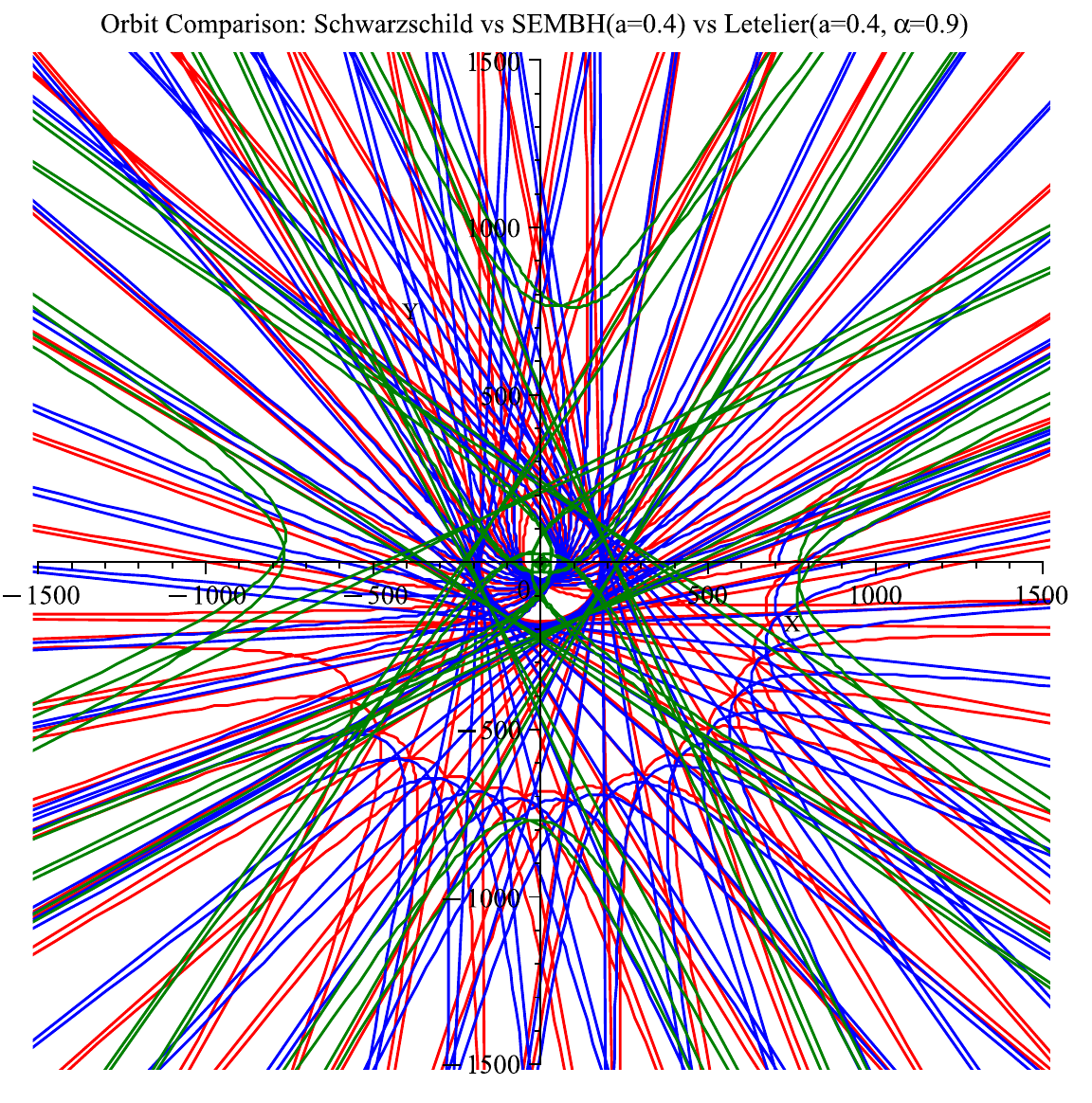}
    \caption{\footnotesize Orbit comparison for fixed EMU parameter $a=0.4$ and varying CoS parameter $\alpha$. Panel (a): $\alpha=0.05$ (weak CoS); Panel (b): $\alpha=0.1$ (moderate CoS); Panel (c): $\alpha=0.9$ (strong CoS). Red curves: SBH reference ($a=1$, $\alpha=0$); blue curves: SEBH with $a=0.4$ and $\alpha=0$; green curves: Letelier BH in EMU with $a=0.4$ and specified $\alpha$. The SBH and SEBH references remain unchanged across panels; only the Letelier orbit (green) expands with increasing $\alpha$. Panel (c) reveals extreme outward migration at $\alpha=0.9$, where the orbit becomes vastly extended due to near-dissolution of gravitational binding as $(1-\alpha)\to 0.1$. All panels use $M=1$, $\mathrm{L}=10$.}
    \label{fig:letelier_orbits_varying_alpha}
\end{figure*}

Figures~\ref{fig:letelier_orbits_varying_a} and~\ref{fig:letelier_orbits_varying_alpha} illustrate the relativistic orbital dynamics of test particles in the Letelier BH spacetime immersed in EMU, comparing three distinct geometries under varying parameters.

Figure~\ref{fig:letelier_orbits_varying_a} demonstrates the influence of the EMU parameter $a$ on test particle trajectories through three-way comparison at nearly vanishing CoS ($\alpha=0.01$). For the extremal RN configuration ($a=0$, panel a), strong EM fields dominate the dynamics, producing the tightest orbital configurations. The three spacetime types exhibit maximum hierarchical separation: the SBH reference (red) shows intermediate-scale precessing orbits, the SEBH (blue) displays the innermost trajectories due to EM enhancement of gravitational attraction, and the Letelier configuration (green) lies slightly outward due to the small CoS dilution. The rosette patterns are highly compressed with rapid perihelion advance, indicating strong gravitational confinement from the $(1-a^2)M^2/r^2$ term with $a=0$.

As $a$ increases to 0.5 (panel b), the EM field strength decreases by a factor of $(1-a^2)=0.75$, allowing all orbits to expand. The blue SEBH orbit moves away from its panel (a) position toward the red SBH reference. The rosette patterns become more open with reduced precession rates. Panel (c) demonstrates that at $a=0.95$ (where $1-a^2=0.0975$), all three orbit types nearly coalesce. The blue SEBH orbit almost overlaps the red SBH reference, while the green Letelier orbit shows minimal separation dominated by the $\alpha=0.01$ factor rather than EM effects.

Figure~\ref{fig:letelier_orbits_varying_alpha} illustrates the complementary effect of CoS density on orbital dynamics at fixed moderate EM field ($a=0.4$). The SBH (red) and SEBH (blue) references remain identical across all panels, serving as stable baselines; only the Letelier orbit (green) changes with $\alpha$. At $\alpha=0.05$ (panel a), the green trajectory shows modest outward displacement from the blue SEBH orbit. The orbital size and precession rate remain comparable to the SEBH case, indicating weak CoS influence when $(1-\alpha)=0.95$. As $\alpha$ increases to 0.1 (panel b), the green orbit expands more noticeably with a wider rosette pattern.

Panel (c) reveals the most striking result: at strong CoS ($\alpha=0.9$), the green Letelier orbit expands to vastly larger scales. The orbital structure transforms qualitatively-instead of a tightly precessing rosette, the trajectory exhibits nearly straight-line segments at large radii before gentle gravitational deflection accumulates. This weak-field behavior arises because $(1-\alpha)=0.1$ reduces the effective gravitational constant by 90\%, requiring orbital radii an order of magnitude larger to maintain the same specific angular momentum. The photon sphere at this configuration resides at $r_{\rm ph}=29.4M$ (Table~\ref{tab:photon_sphere_results}), explaining the extended orbital scales.

The comparison between Figs.~\ref{fig:letelier_orbits_varying_a} and~\ref{fig:letelier_orbits_varying_alpha} reveals fundamentally different mechanisms by which EM fields and CoS modify orbital dynamics. The EM parameter $a$ affects all three curves through different channels and produces smooth, monotonic convergence toward a common SBH-like limit as $a\to 1$, with typical size variations of 30--50\% across the range $a\in[0,1]$. In contrast, the CoS parameter $\alpha$ exclusively affects the Letelier orbit (green) while leaving SBH and SEBH references unchanged, demonstrating its role as an independent global geometric deformation. The $\alpha$-dependence produces dramatic nonlinear scaling: weak CoS ($\alpha<0.2$) causes modest $\sim$20\% orbital expansions, but strong CoS ($\alpha>0.8$) triggers order-of-magnitude size increases.

These distinct parametric behaviors provide complementary observational handles. The EM parameter $a$ primarily affects orbital precession rates and perihelion distances, while the CoS parameter $\alpha$ controls absolute orbital scales. Precision astrometry of stellar orbits around supermassive BHs like Sgr~A* could potentially constrain both parameters: $a$ via perihelion precession measurements \cite{sec3is14} and $\alpha$ via overall orbital size calibration against predicted masses. Gravitational wave observations of extreme mass ratio inspirals would similarly encode both effects, with $a$ modulating high-frequency orbital phase evolution and $\alpha$ affecting the overall inspiral timescale through its influence on binding energy \cite{sec3is15}.

\section{Weak Gravitational Lensing} \label{sec4}

Gravitational lensing serves as one of the most powerful probes of spacetime geometry, offering direct observational access to the mass distribution and metric structure around compact objects \cite{sec4is01,sec4is02}. In the weak-field regime, light rays passing far from the gravitating source experience small but measurable deflections that encode information about the source parameters. In this section, we derive the deflection angle for massive particles and photons in the Letelier BH immersed in EMU using the Jacobi-Maupertuis-Randers-Finsler (JMRF) metric formalism combined with the Gauss-Bonnet theorem \cite{isz41,isz42}. We examine how the CoS parameter $\alpha$ and EMU parameter $a$ modify the lensing signatures compared to the SBH case.

For the static spherically symmetric metric~(\ref{aa4}), the Jacobi-Maupertuis metric on the spatial manifold takes the form \cite{sec4is05}
\begin{equation}
ds_J^2 = \bar{\alpha}_{ij}dx^i dx^j = \left(\frac{E^2}{f(r)} - m^2\right)\left[\frac{dr^2}{f(r)} + r^2(d\theta^2 + \sin^2\theta\, d\phi^2)\right],\label{ee1}
\end{equation}
where $E$ is the conserved energy and $m$ is the rest mass of the test particle. On the equatorial plane ($\theta = \pi/2$), this reduces to
\begin{equation}
\bar{\alpha}_{ij}dx^i dx^j = \left(\frac{E^2}{f(r)} - m^2\right)\left[\frac{dr^2}{f(r)} + r^2 d\phi^2\right].\label{ee2}
\end{equation}

For particles with asymptotic velocity $v$, the energy-velocity relation $E = m/\sqrt{1-v^2}$ yields
\begin{equation}
\bar{\alpha}_{ij}dx^i dx^j = \frac{m^2}{1-v^2}\left[\frac{1-v^2f(r)}{f(r)}\right]\left[\frac{dr^2}{f(r)} + r^2 d\phi^2\right].\label{ee3}
\end{equation}

Following the approach of Werner \cite{isz42}, we construct an osculating Riemannian manifold to apply the Gauss-Bonnet theorem. The first-order particle trajectory in the weak-field limit is
\begin{equation}
r(\phi) = \frac{b}{\sin\phi} - \left(\cot^2\phi + \frac{\csc^2\phi}{v^2}\right)M_{\text{eff}} + O(M_{\text{eff}}^2),\label{ee4}
\end{equation}
where $b$ is the impact parameter. A notable feature of the Letelier geometry is the emergence of an effective mass parameter
\begin{equation}
M_{\text{eff}} = \frac{M}{1-\alpha},\label{ee5}
\end{equation}
which shows that CoS matter effectively enhances the gravitational mass by a factor $(1-\alpha)^{-1}$. This amplification arises from the global $(1-\alpha)$ factor in the metric function~(\ref{aa5}) and has direct observational consequences for lensing measurements.

The vector fields along the zero-order trajectory are
\begin{equation}
Y^r = -\frac{\cos\phi}{mv}\sqrt{1-v^2}, \quad Y^\phi = \frac{\sin^2\phi}{mbv}\sqrt{1-v^2}.\label{ee6}
\end{equation}

The osculating Riemannian metric components, obtained from the Hessian of the Finsler metric, are
\begin{align}
\tilde{g}_{rr} &= \frac{m^2}{1-v^2}\bigg[v^2 + \frac{2M(1+v^2)}{r(1-\alpha)} - \frac{2M^2(1-a^2)(1+v^2)}{r^2(1-\alpha)}\nonumber\\
&\quad + \frac{4M^2(2+v^2)}{r^2(1-\alpha)^2}\bigg] + O(\epsilon^3),\label{ee7}
\end{align}
\begin{align}
\tilde{g}_{\phi\phi} &= \frac{m^2r^2}{1-v^2}\bigg[v^2 + \frac{2M}{r(1-\alpha)} - \frac{2M^2(1-a^2)}{r^2(1-\alpha)}\nonumber\\
&\quad + \frac{4M^2}{r^2(1-\alpha)^2}\bigg] + O(\epsilon^3),\label{ee8}
\end{align}
where $\epsilon^3$ denotes third-order terms in $M$, $\alpha$, and $(1-a^2)$.

\subsection{Gaussian Curvature and Deflection Angle}

The Gaussian curvature of the osculating Riemannian manifold is computed using the standard formula \cite{sec4is07}
\begin{equation}
\tilde{K} = \frac{1}{\sqrt{\det\tilde{g}}}\left[\frac{\partial}{\partial\phi}\left(\frac{\sqrt{\det\tilde{g}}}{\tilde{g}_{rr}}\Gamma^{\phi}_{rr}\right) - \frac{\partial}{\partial r}\left(\frac{\sqrt{\det\tilde{g}}}{\tilde{g}_{rr}}\Gamma^{\phi}_{r\phi}\right)\right].\label{ee9}
\end{equation}

After detailed calculation, the Gaussian curvature becomes
\begin{align}
\tilde{K} &= -\frac{(1-v^4)M}{m^2r^3v^4(1-\alpha)} + \frac{(1-v^4)M^2(1-a^2)}{m^2r^4v^4(1-\alpha)} \nonumber\\
&\quad + \frac{3(2-3v^2+v^4)M^2}{m^2r^4v^6(1-\alpha)^2} + O(\epsilon^3).\label{ee10}
\end{align}
The first term represents the leading-order curvature enhanced by the CoS factor $(1-\alpha)^{-1}$. The second term contains the EMU correction proportional to $(1-a^2)$, which vanishes in the SBH limit $a\to 1$. The third term provides the pure CoS correction at second order.

Applying the Gauss-Bonnet theorem to the region outside the lens yields \cite{sec4is08}
\begin{equation}
\hat{\alpha}_{|\infty} = -\iint_{D_\infty} \tilde{K}\,dS = -\int_0^\pi d\phi \int_{b/\sin\phi}^\infty \tilde{K}\sqrt{\det\tilde{g}} \, dr.\label{ee11}
\end{equation}

Performing the integration yields the infinite-distance deflection angle for massive particles:
\begin{align}
\hat{\alpha}_{|\infty} &= \frac{2M(1+v^2)}{bv^2(1-\alpha)} - \frac{\pi M^2(1-a^2)(1+v^2)}{2b^2v^2(1-\alpha)} \nonumber\\
&\quad + \frac{3\pi(4+v^2)M^2}{4b^2v^2(1-\alpha)^2} + O(\epsilon^3).\label{ee12}
\end{align}

For photons ($v = 1$), this reduces to
\begin{equation}
\hat{\alpha}_{|\infty}^{\text{light}} = \frac{4M}{b(1-\alpha)} - \frac{\pi M^2(1-a^2)}{b^2(1-\alpha)} + \frac{15\pi M^2}{4b^2(1-\alpha)^2} + O(\epsilon^3).
\label{ee13}
\end{equation}

\subsection{Physical Interpretation and Limiting Cases}

The deflection angle formula~(\ref{ee13}) reveals the interplay between CoS and EMU effects on gravitational lensing. The first term gives the primary deflection, enhanced by the factor $(1-\alpha)^{-1}$ due to the gravitational amplification by CoS matter. This leading-order term scales as $M/b$, the familiar weak-field result, but with an effective mass $M_{\rm eff}=M/(1-\alpha)$ that exceeds the bare mass for any $\alpha>0$. The second term, proportional to $(1-a^2)$, represents the EMU correction at second order. Notably, this term carries a negative sign, indicating that the EM background \textit{reduces} the deflection angle for $a<1$. This reduction can be understood physically: the $(1-a^2)M^2/r^2$ term in the metric function~(\ref{aa5}) acts as an attractive potential that pulls light rays closer to straight-line trajectories at large distances. The third term provides the pure CoS correction at second order, enhanced by $(1-\alpha)^{-2}$.

Several limiting cases illuminate the physical content of Eq.~(\ref{ee13}):

(i) \textit{SBH limit} ($a=1$, $\alpha=0$): The deflection angle reduces to
\begin{equation}
\hat{\alpha}_{|\infty} = \frac{4M}{b} + \frac{15\pi M^2}{4b^2},\label{ee14}
\end{equation}
recovering the standard SBH result with second-order correction \cite{sec4is09}.

(ii) \textit{Letelier BH} ($a=1$, $\alpha\neq 0$): The deflection becomes
\begin{equation}
\hat{\alpha}_{|\infty} = \frac{4M}{b(1-\alpha)} + \frac{15\pi M^2}{4b^2(1-\alpha)^2},\label{ee15}
\end{equation}
showing the pure CoS enhancement without EMU corrections \cite{sec4is10}.

(iii) \textit{SEBH} ($\alpha=0$, $a\neq 1$): The deflection is
\begin{equation}
\hat{\alpha}_{|\infty} = \frac{4M}{b} - \frac{\pi M^2(1-a^2)}{b^2} + \frac{15\pi M^2}{4b^2},\label{ee16}
\end{equation}
revealing the EMU reduction at second order.

(iv) \textit{Extremal RN limit} ($a=0$, $\alpha=0$): The deflection becomes
\begin{equation}
\hat{\alpha}_{|\infty} = \frac{4M}{b} - \frac{\pi M^2}{b^2} + \frac{15\pi M^2}{4b^2} = \frac{4M}{b} + \frac{11\pi M^2}{4b^2},\label{ee17}
\end{equation}
where the EMU and geometric corrections partially cancel.

\begin{figure*}[ht!]
\centering
\includegraphics[width=0.85\textwidth]{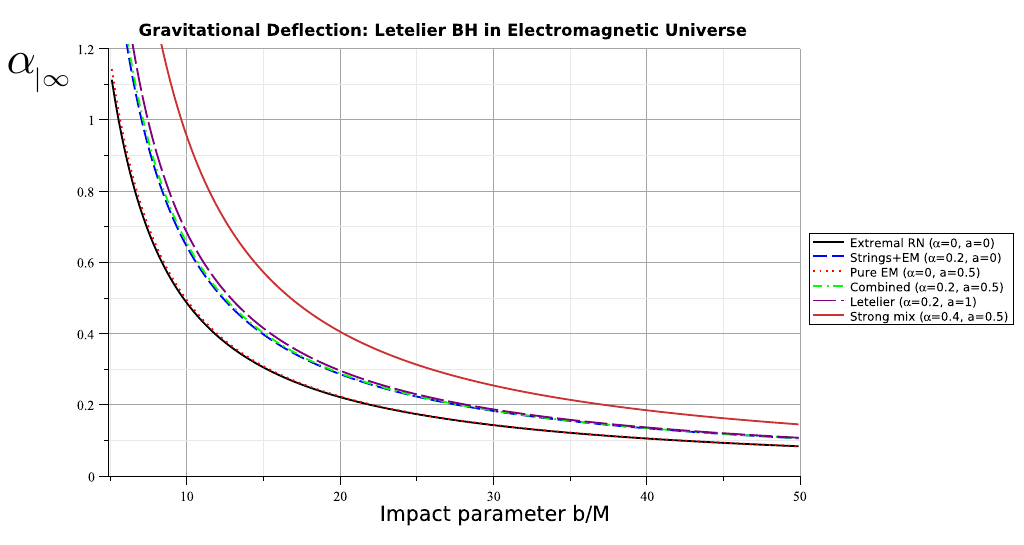}
\caption{\footnotesize Deflection angle versus impact parameter for various combinations of CoS parameter $\alpha$ and EMU parameter $a$. The curves demonstrate the competing effects: CoS enhances deflection through the $(1-\alpha)^{-1}$ factor, while the EM field reduces it via the $(1-a^2)$ term at second order. Mass $M=1$ throughout.}
\label{fig:deflection_curves}
\end{figure*}

\begin{figure}[ht!]
\centering
\includegraphics[width=0.5\textwidth]{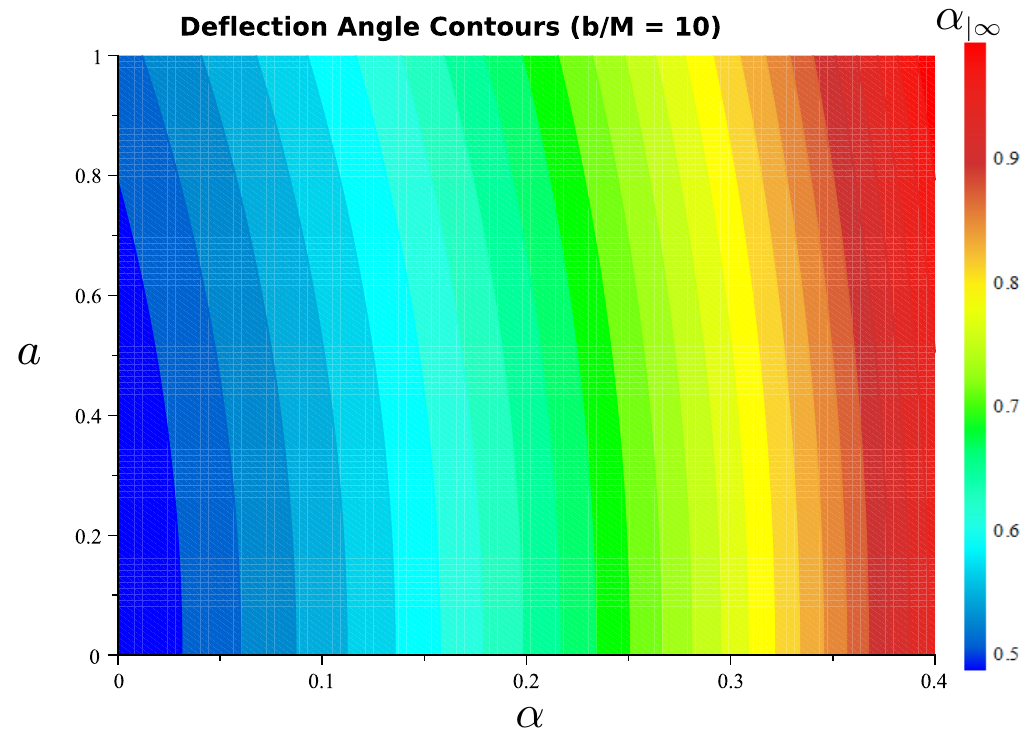}
\caption{\footnotesize Parameter space exploration showing deflection angle behavior for different EMU strengths. The variation reveals how the EM parameter $a$ modulates the CoS enhancement, with maximum deflection at $a=0$ (extremal RN limit) and minimum at $a=1$ (SBH limit). Mass $M=1$.}
\label{fig:deflection_em_effect}
\end{figure}

\begin{figure}[ht!]
\centering
\includegraphics[width=0.5\textwidth]{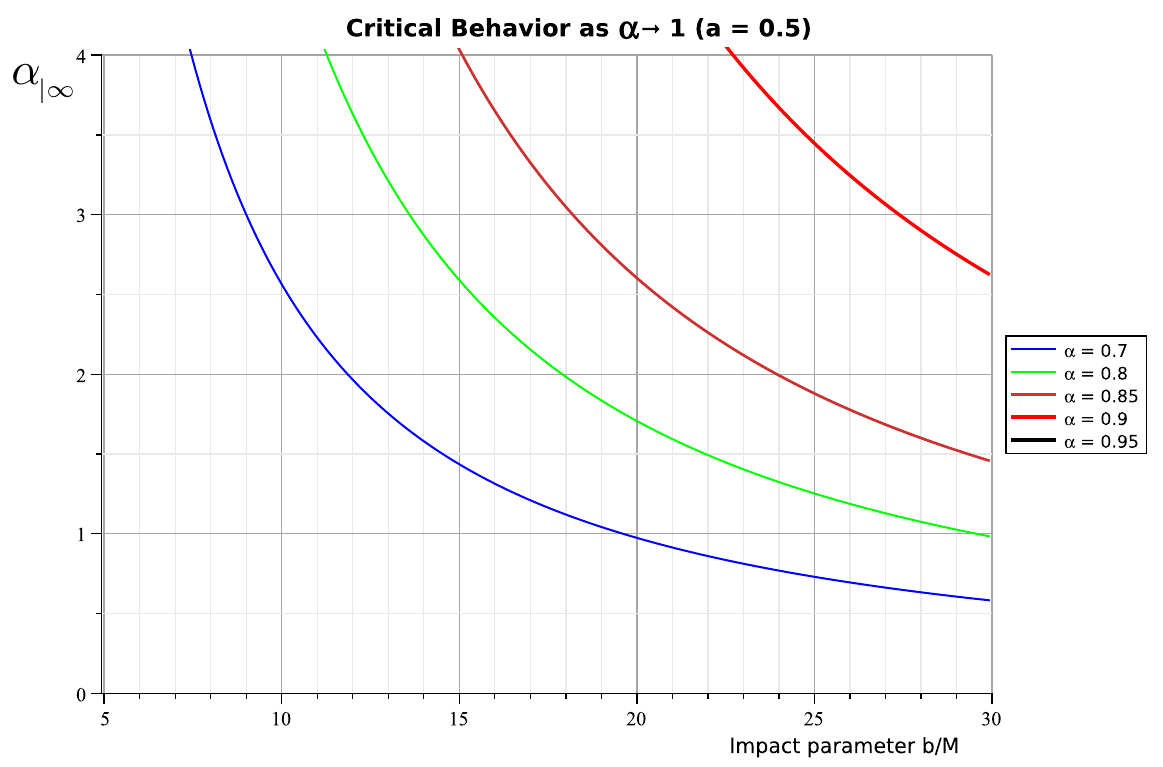}
\caption{\footnotesize Contour plot of the deflection angle in the $(\alpha, a)$ parameter space for fixed impact parameter $b/M = 10$. The gradient from blue (low deflection) to red (high deflection) reveals the nonlinear interplay between CoS and EMU effects, with maximum deflection in the high-$\alpha$, low-$a$ region. The forbidden region near $\alpha\to 1$ corresponds to naked singularity formation.}
\label{fig:deflection_contour}
\end{figure}

Figures~\ref{fig:deflection_curves}--\ref{fig:deflection_contour} visualize the lensing phenomenology in the Letelier BH immersed in EMU. Figure~\ref{fig:deflection_curves} demonstrates the competition between CoS enhancement and EMU reduction. The CoS amplifies deflection through the $(1-\alpha)^{-1}$ factor in the leading term, while the EMU introduces a reduction proportional to $(1-a^2)$ at second order. This creates a rich spectrum of deflection behaviors, with the most dramatic enhancement occurring for large $\alpha$ and small $a$ values. The curves show that CoS effects dominate at small impact parameters where the $(1-\alpha)^{-1}$ factor has maximum influence, while EMU corrections become more significant at intermediate impact parameters through the second-order terms.

Figure~\ref{fig:deflection_em_effect} explores the parameter space in more detail, revealing that the EMU parameter $a$ acts as a modulating factor. When $a\to 0$ (approaching the extremal RN limit), the EM field maximally reduces the second-order deflection, while as $a\to 1$ (SBH limit), this reduction vanishes. However, the CoS enhancement in the leading term remains independent of $a$, ensuring that large $\alpha$ values always produce significant deflection amplification regardless of the EM background strength.

The contour plot in Fig.~\ref{fig:deflection_contour} synthesizes these effects, showing a gradient from the lower-left corner (minimal deflection for small $\alpha$ and large $a$) to the upper-left region (maximal deflection for large $\alpha$ and small $a$). The approximately diagonal contours indicate partial degeneracy: the two parameters can compensate for each other to maintain constant deflection. Increasing the EM field strength (decreasing $a$) can offset a reduction in CoS density (decreasing $\alpha$) within certain ranges. This degeneracy could complicate observational constraints but also provides multiple parameter combinations that could explain observed lensing signatures.

The forbidden region approaching $\alpha\to 1$ corresponds to naked singularity formation (cf. Table~\ref{izzetletelier_horizons}) and represents a physical boundary of the parameter space. These distinctive lensing signatures offer potential observational tests to distinguish Letelier BHs in EMU from standard BH solutions, particularly through precision astrometric measurements of stellar deflections near galactic centers \cite{sec4is11} or detailed analysis of gravitational lensing arcs in strong lensing systems \cite{sec4is12}.

\section{Conclusion}\label{sec5}

In this work, we investigated the physical properties of Letelier BHs immersed in an EMU, a configuration that combines the gravitational effects of a CoS with the EM background characteristic of the BR spacetime. The resulting geometry, described by the metric function $f(r) = 1 - \alpha - 2M/r + M^2(1-a^2)/r^2$ in Eq.~(\ref{aa5}), interpolates between several well-known solutions: the SBH when $\alpha=0$ and $a=1$, the Letelier BH when $a=1$, the SEBH when $\alpha=0$, and the extremal RN geometry when $a=0$ and $\alpha=0$. The two parameters $\alpha$ (CoS density) and $a$ (EMU parameter) encode fundamentally different physical mechanisms-global gravitational dilution and local EM confinement, respectively-whose interplay produced the rich phenomenology explored throughout this study \cite{isz01,isz02}.

In Sec.~\ref{isec2}, we derived the horizon structure of the Letelier BH immersed in EMU. The event horizon radii, given by Eq.~(\ref{aa7}), revealed that real horizons exist when the condition $(1-\alpha)(1-a^2) \leq 1$ is satisfied. Table~\ref{izzetletelier_horizons} and Fig.~\ref{izzetletelier_metric_function} demonstrated that increasing $\alpha$ toward unity eventually eliminates the event horizon, producing a naked singularity. The CoS parameter entered through the $(1-\alpha)^{-1}$ prefactor, which amplified the horizon radius as $\alpha$ increased, while the EMU parameter appeared in the discriminant, controlling the separation between inner and outer horizons. In the SBH limit, we recovered $r_+ = 2M$, validating our analytical expressions.

The thermodynamic analysis in Sec.~\ref{sec3a} uncovered several notable features. The Hawking temperature, derived in Eq.~(\ref{ss4}), required the normalization factor $C = (1-\alpha)^{-1/2}$ to ensure proper behavior in the asymptotically bounded spacetime created by CoS matter. The heat capacity, given by Eq.~(\ref{ss7}), exhibited a divergence at the critical radius $r_c = \sqrt{3(1-a^2)/(1-\alpha)}\,M$, signaling a second-order phase transition between thermodynamically stable (small BHs with $C > 0$) and unstable (large BHs with $C < 0$) configurations \cite{isz03,isz04}. Figure~\ref{fig:surface-gravity} illustrated how the surface gravity varied with the EMU parameter $a$ for different CoS densities, showing that both parameters influenced the thermal properties in distinct ways.

Section~\ref{sec3b} addressed the dynamics of massive test particles. The effective potential for timelike geodesics, Eq.~(\ref{aa11}), exhibited an asymptotic value $V_{\rm eff} \to 1-\alpha$ as $r \to \infty$, directly reflecting the gravitational dilution by CoS matter. Figure~\ref{fig:time-like-potential} demonstrated that increasing $\alpha$ lowered the potential barrier and reduced the spacetime's binding capacity, while decreasing $a$ (stronger EM fields) enhanced gravitational confinement near the horizon. The ISCO radius, determined by the cubic equation~(\ref{aa14}), increased monotonically with both $\alpha$ and $a$, as visualized in the three-dimensional plot of Fig.~\ref{fig:ISCO}. In the SBH limit, we recovered the familiar result $r_{\rm ISCO} = 6M$.

The stability of the Letelier BH immersed in EMU was examined through scalar and EM perturbations in Secs.~\ref{sec3c} and~\ref{sec3d}. The effective potential for scalar perturbations, Eq.~(\ref{bb3}), governed the wave equation in tortoise coordinates. We computed the QNM frequencies using the sixth-order WKB approximation, with results presented in Tables~\ref{taba3} and~\ref{taba4}. The imaginary parts of all computed frequencies remained negative, confirming the linear stability of the BH under both scalar and EM perturbations \cite{isz05,isz06}. The real part of the QNM frequency (oscillation rate) decreased with both $\alpha$ and $a$, reflecting the expansion of the photon sphere with these parameters. The CoS parameter produced a more dramatic effect than the EMU parameter: as $\alpha$ increased from $0.1$ to $0.45$, the oscillation frequency dropped by more than $50\%$, compared to $\sim 15\%$ variation across the $a$ range studied. Figures~\ref{fig:placeholder}--\ref{fig:placeholder3} visualized these trends for both scalar and EM perturbations.

In Sec.~\ref{sec3e}, we analyzed the photon sphere and null geodesic structure. The photon sphere radius, given analytically by Eq.~(\ref{dd2}), demonstrated the opposing effects of the two parameters: the CoS expanded the photon sphere through the $(1-\alpha)^{-1}$ factor, while the EMU contracted it through the $(1-a^2)$ term under the square root. Table~\ref{tab:photon-sphere} provided a parametric survey confirming these trends, with $r_{\rm ph}$ ranging from $2M$ at the extremal RN limit to $4M$ at $(\alpha=0.25, a=1)$. The BH shadow radius, Eq.~(\ref{shadow}), encoded both parameters and offered a potential observational probe through EHT-type measurements. Figures~\ref{fig:letelier_orbits_varying_a} and~\ref{fig:letelier_orbits_varying_alpha} illustrated the orbital dynamics, revealing that strong CoS ($\alpha = 0.9$) produced order-of-magnitude expansions in orbital scales due to near-dissolution of gravitational binding.

Section~\ref{sec4} derived the weak-field gravitational lensing signatures using the JMRF formalism and Gauss-Bonnet theorem. The deflection angle for photons, Eq.~(\ref{ee13}), exhibited three distinct contributions: the leading-order term enhanced by $(1-\alpha)^{-1}$, the second-order EMU correction proportional to $-(1-a^2)$, and the pure CoS correction scaling as $(1-\alpha)^{-2}$. The CoS enhancement and EMU reduction created competing effects, with the former dominating at small impact parameters and the latter becoming significant at intermediate distances. Figures~\ref{fig:deflection_curves}--\ref{fig:deflection_contour} visualized this interplay, showing that maximum deflection occurred in the high-$\alpha$, low-$a$ region of parameter space. The contour plot revealed approximate degeneracies where different parameter combinations produced identical deflection angles, which could complicate observational constraints but also provide flexibility in fitting observed lensing data.

Several directions for future research emerge from this study. First, extending the analysis to rotating Letelier BHs in EMU would capture spin effects relevant for astrophysical BHs and enable comparison with Kerr observations. Second, computing greybody factors and absorption cross-sections would complement the QNM analysis and provide additional observational signatures. Third, investigating the strong-field lensing regime, including relativistic images and photon rings, would probe the near-horizon geometry more directly. Fourth, exploring the thermodynamic topology and studying the Ruppeiner geometry of the phase space could reveal microscopic structures underlying the observed phase transitions. Fifth, analyzing gravitational wave emission from extreme mass ratio inspirals in this background would connect the theoretical predictions to LISA-band observations. Finally, constraining the parameters $\alpha$ and $a$ using EHT shadow measurements and stellar orbit data from the Galactic Center represents a promising observational avenue that could test the viability of CoS and EMU modifications to standard BH physics.

In summary, the Letelier BH immersed in EMU exhibits a rich phenomenology arising from the interplay between CoS gravitational dilution and EMU EM confinement. The two parameters $\alpha$ and $a$ control distinct aspects of the geometry: $\alpha$ governs global properties such as asymptotic behavior and overall scales, while $a$ modifies local structure near the horizon. The stability under perturbations, distinctive thermodynamic phase structure, modified ISCO and photon sphere locations, and characteristic lensing signatures collectively provide a framework for testing this class of solutions against astrophysical observations.

\section*{Acknowledgment}

F.A. gratefully acknowledges the Inter University Centre for Astronomy and Astrophysics (IUCAA), Pune, India, for the conferment of a visiting associateship. \.{I}.~S. thanks T\"{U}B\.{I}TAK, ANKOS, and SCOAP3 for academic support and acknowledges the networking support from COST Actions CA22113, CA21106, CA23130, CA21136, and CA23115. 

\section*{Data Availability Statement}

No new data were generated or analyzed in this manuscript.

\section*{Conflict of Interest}

Authors declare(s) no conflict of interest.

\end{document}